\documentclass[12pt]{article}
\pdfoutput=1
\usepackage{amsmath,amssymb,bm,cite,graphicx,slashed,subcaption,url}
\usepackage[T1]{fontenc}

\usepackage{amsfonts,fullpage}
\usepackage[font=normal, width=0.90\linewidth]{caption}

\usepackage{color}
\definecolor{colorLink}{rgb}{0.9,0,0} % red
\definecolor{colorCite}{rgb}{0,0.7,0} % green
\definecolor{colorURL} {rgb}{0,0,0.8} % navy
\definecolor{colorCapt}{rgb}{0,0.7,0} % green
\usepackage[colorlinks=true,linktocpage=true,linkcolor=colorLink,citecolor=colorCite,urlcolor=colorURL]{hyperref}
\usepackage[labelfont={color=colorCapt,bf}]{caption}

\newcommand{\cO}{\mathcal{O}}
\newcommand{\cN}{\mathcal{N}}

\definecolor{colorRTD}{rgb}{.2,.2,.7}

\definecolor{colorML}{rgb}{.2,.7,.7}

\usepackage{tikz}
\usetikzlibrary{arrows,shapes}
\usetikzlibrary{trees}
\usetikzlibrary{matrix,arrows}
\usetikzlibrary{positioning}
\usetikzlibrary{calc,through}
\usetikzlibrary{intersections}
\usetikzlibrary{decorations.pathreplacing}
\usetikzlibrary{decorations.pathmorphing}
\usetikzlibrary{decorations.markings}
\usepackage{pgffor}
\usetikzlibrary{tikzmark,calc}
\tikzset{
    vector/.style={decorate, decoration={snake}, draw},
    provector/.style={decorate, decoration={snake,amplitude=2.5pt}, draw},
    antivector/.style={decorate, decoration={snake,amplitude=-2.5pt}, draw},
    fermion/.style={draw=black, postaction={decorate},decoration={markings,mark=at position .55 with {\arrow[draw=black]{>}}}},
    fermionbar/.style={draw=black, postaction={decorate},decoration={markings,mark=at position .55 with {\arrow[draw=black]{<}}}},
    fermionnoarrow/.style={draw=black},
    doublefermionnoarrow/.style={double,draw=black},
    gluon/.style={decorate, draw=black,decoration={coil,amplitude=4pt, segment length=5pt}},
    scalar/.style={dashed,draw=black, postaction={decorate},decoration={markings,mark=at position .55 with {\arrow[draw=black]{>}}}},
    scalarbar/.style={dashed,draw=black, postaction={decorate},decoration={markings,mark=at position .55 with {\arrow[draw=black]{<}}}},
    scalarnoarrow/.style={dashed,draw=black},
    doublescalarnoarrow/.style={dashed,double,draw=black},
    electron/.style={draw=black, postaction={decorate},decoration={markings,mark=at position .55 with {\arrow[draw=black]{>}}}},
    bigvector/.style={decorate, decoration={snake,amplitude=4pt}, draw},
}
\title{
\vspace{-2cm}
\begin{flushright}{\small FERMILAB-PUB-18-687-T}\vspace{2cm}\end{flushright}
\bf{\Large Disorder and Mimesis at Hadron Colliders}}
\date{\today}

\author{Raffaele Tito D'Agnolo$^{1}$ and Matthew Low$^{2,3}$\\[2pt]
{\small\emph{$^{1}$ {SLAC National Accelerator Laboratory, 2575 Sand Hill Road, Menlo Park, CA, 94025, USA.}}}\\
{\small\emph{$^{2}$School of Natural Sciences, Institute for Advanced Study, Princeton, NJ 08540}}\\
{\small\emph{$^{3}$Theoretical Physics Department, Fermilab, P.O. Box 500, Batavia, IL 60510}}}

\begin{document}
\maketitle

%--------------------------------------------- ABSTRACT ---------------------------------------------%

\abstract{We discuss how systems with a large number of degrees of freedom and disorder in their mass matrix can play a role in particle physics.  We derive results on their mass spectra using, where applicable, QFT techniques.  We study concrete realizations of these scenarios in the context of the LHC and HL-LHC, showing that collider events with a large number of soft $b$-quark jets can be common.  Such final states can hide these models from current searches at the LHC.  This motivates the ongoing effort aimed at lowering trigger thresholds and expanding data scouting.}

\newpage
\tableofcontents

%**************** Section ***********************************
\section{Introduction}
\label{sec:intro}
%*************************************************************

The current progress of the experimental effort at the Large Hadron Collider (LHC) has largely exceeded expectations.  An unprecedented amount of high-quality data has been collected and true milestones have been reached for the field, such as the discovery of the Higgs boson~\cite{Aad:2012tfa,Chatrchyan:2012xdj}.  After ten years of operation, however, and hundreds of measurements that constrain all the most plausible scenarios for physics beyond the Standard Model (SM), we are left to wonder if there are any new particles hiding at the weak scale. 

This is the right time to ask this question because we have already a rather extensive picture of physics around a TeV from the studies performed on tens of fb$^{-1}$ of data at $\sqrt{s}=13~{\rm TeV}$~\cite{ATLASResults,CMSResults}.  The upcoming years of LHC operation will be characterized by a very different pace, determined by a slow increase in sensitivity driven by the collected integrated luminosity.

This question has inspired a large part of the theoretical effort for the past several years, mainly in the direction of finding new solutions to the hierarchy problem hidden from traditional LHC searches.  By now most of these new scenarios are significantly constrained.  Here we would like to take a completely different perspective and abandon all prior theoretical expectations.  If we do so, two possible answers strike us as the simplest and potentially most plausible: either we have found nothing beyond the Higgs boson because (1) there is nothing to be found at the weak scale\footnote{Including the possibility that new physics is too weakly-coupled for us to detect it.} or (2) there are too many new particles. 

In this paper we show how large $N$ sectors are naturally hard to detect at hadron colliders.  The reasons are simple and independent of a specific model.  The first one is that they require a small coupling to be consistent. This together with the finite kinematical range accessible to colliders can give a small total production rate. The second reason is that high-multiplicity final states containing only low $p_T$ particles can easily dominate their total production rate, presenting a challenge for current triggers.  We discuss this in Sect.~\ref{sec:kinematics}.

Emphasizing this general point about large $N$ sectors is useful both from the experimental and theoretical perspective.  On the experimental side, anything that can be lost due to current triggers deserves serious consideration. Missing new physics because of our own choices in the selection of events would be a highly tragic mistake. An extensive effort in this direction is already in progress at the LHC, in the form of data parking and data scouting~\cite{CMS:2012ooa, Mukherjee:2017wcl, Duarte:2018bsd, Anderson:2016ron} and upgrades for the high-luminosity run, such as implementing particle flow at Level 1~\cite{Kreis:2018job, petruccianiL1} in CMS and the Feature Extractors~\cite{Stark:2016fnm, Intent} and Fast Track Trigger (FTK)~\cite{Shochet:2013gaw} in ATLAS.

On the theory side, even if the mimetic properties of large $N$ sectors are not inspired by any open theoretical question, the models that realize them are well-motivated.  They can arise as remnants of string theory compactifications~\cite{Halverson:2018xge, Cvetic:2002qa} and/or as a low energy sector of the landscape~\cite{ArkaniHamed:2005yv}.  They can also be part of a hidden sector containing dark matter or modifying the electroweak phase transition, giving rise to a phase of symmetry non-restoration~\cite{Meade:2018saz,Baldes:2018nel,Glioti:2018roy}.  They are related to the broader framework of hidden valleys~\cite{Strassler:2006im, Han:2007ae, Strassler:2008fv, Strassler:2006ri} and realize a phenomenology that is in-between that of traditional hidden valleys (a small number of possibly displaced SM particles in the final state) and that of conformal hidden valleys~\cite{Strassler:2008bv,Knapen:2016hky} (many particles emitted isotropically). 

Furthermore, from a bottom-up viewpoint, the large $N$ sectors that we discuss offer the perfect opportunity to study the possible role of disorder in model building.\footnote{See~\cite{Dienes:2016kgc} for uses in dynamical dark matter~\cite{Dienes:2011ja,Dienes:2011sa}.} In our construction disorder is nothing more than a useful phenomenological tool. It allows us to capture possible $\cO(1)$ variations of the low energy parameters of the theory. However,  as it will emerge in the following, disordered systems possess interesting structural properties and in the future their significance in beyond the SM (BSM) physics might be much greater than this. Therefore we use this opportunity to discuss a number of results on random matrices in a heuristic way, useful for model building. In the appendices we expand our derivations, making them more rigorous and using, where applicable, path integral techniques familiar from QFT.

The paper is organized as follows: in Sect.~\ref{sec:kinematics} we discuss the general kinematical mechanism that hides large $N$ sectors from detection. In Sect.~\ref{sec:models} we present a concrete realization of the ideas discussed in  Sect.~\ref{sec:kinematics}, in the form of a disordered, large $N$ model of scalars. In Sect.~\ref{sec:disorder} we discuss general results on disordered mass matrices of scalars, in Sect.~\ref{sec:pheno} we study the collider phenomenology of these models, and in Sect.~\ref{sec:fermions} we discuss large $N$ models containing fermions.  We conclude in Sect.~\ref{sec:outlook} suggesting next steps to further develop this framework.

%**************** Section ***********************************
\section{Large $N$ Mimesis}
\label{sec:kinematics}
%*************************************************************

The reasons that make a new sector with a large number of new particles hard to detect at hadron colliders are very simple and completely general.  We present them briefly in this section before discussing a specific model.  Here we imagine that the number $N$ of particles is always large, but the kinematic arguments still apply to a moderate number of particles, even as few as 5.

%%%%%%%%%%%%%%%%%%%%%%%%%%%%%%%%%%%%%%%
\begin{figure}[!t]
\centering
  \begin{tikzpicture}[scale=1.6]
  \draw[->,line width=0.8pt] (-0.5 ,-0.8) -- (-0.5 , 1.8);
  \node[rotate=90] at (-0.6, 0.5 ) {mass};
  %%%
  \filldraw[draw=none,fill=lightgray, fill opacity=0.6, rounded corners=2pt] (-0.1 ,-0.1 ) rectangle ( 2.1 , 1.7 ) ;
  \node at ( 1.0, 1.4) {$\vdots$};
  \draw (  0,0.9) -- (2.0,0.9);
  \draw (  0,0.8) -- (2.0,0.8);
  \draw (  0,0.7) -- (2.0,0.7);
  \draw (  0,0.6) -- (2.0,0.6);
  \draw (  0,0.5) -- (2.0,0.5);
  \draw (  0,0.4) -- (2.0,0.4);
  \draw (  0,0.3) -- (2.0,0.3);
  \draw (  0,0.2) -- (2.0,0.2);
  \draw (  0,0.1) -- (2.0,0.1);
  \draw (  0,0.0) -- (2.0,0.0);
  \draw (0,-0.4) -- (2.0,-0.4);
  \draw (0,-0.6) -- (2.0,-0.6);
  \draw (0,-0.7) -- (2.0,-0.7);
  %%%
  \draw [decorate,decoration={brace,amplitude=5pt,mirror,raise=4pt},yshift=0pt,line width=0.8pt] (2.2,-0.8) -- (2.2,-0.3) 
        node [black,midway,xshift=0.4cm,right] {\footnotesize accessible ($\sim N_{\rm LHC}$)};
  \draw [decorate,decoration={brace,amplitude=5pt,mirror,raise=4pt},yshift=0pt,line width=0.8pt] (2.2,-0.1) -- (2.2, 1.7) 
        node [black,midway,xshift=0.4cm,right] {\footnotesize inaccessible ($\sim N$)};
  \end{tikzpicture}
\captionsetup{width=0.87\linewidth}
\caption{A large $N$ theory where only $N_{\rm LHC} \ll N$ states are kinematically accessible to a collider gives a production cross section that is suppressed by powers of $N_{\rm LHC}/N$.}
\label{fig:scheme1}
\end{figure}
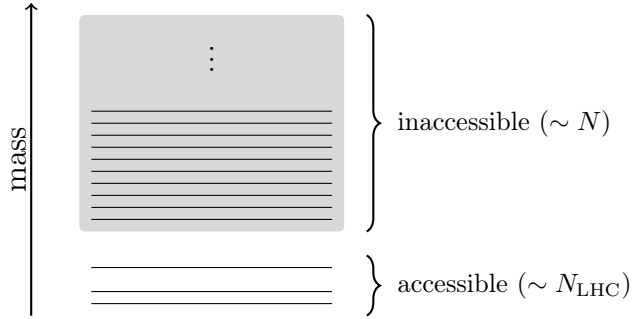
%%%%%%%%%%%%%%%%%%%%%%%%%%%%%%%%%%%%%%%

Introducing $N\gg 1$ particles in a finite mass range has three main consequences relevant for colliders:
\begin{enumerate}
\item The spectrum is compressed and final states with only soft particles are common.
\item The theory requires a small coupling to be consistent.
\item Long decay chains can arise naturally and dominate the total production rate.
\end{enumerate}
The first two items tend to make new sectors with a relatively large number of new particles hard to detect.  The last one can be used as a handle to disentangle these new sectors from the background and even reconstruct their structure.  However, long decay chains and soft final states go hand-in-hand so having more particles in the final state is not necessarily advantageous.  Let us now discuss each of these three aspects in more detail. 

Compression is a trivial consequence of having a large number of new states in a finite mass range. However the amount of compression depends only on the density of particles per unit mass, so it can persist also with a small number of new particles.  

The small coupling arises if we require these theories to be perturbatively consistent.  Diagrams that are typically higher-order in the couplings, like loop diagrams, involve sums over the new particles and when their number $N$ is large the couplings must compensate by scaling with the appropriate power of $1/N$.  In this regime it is useful to reorganize the perturbative expansion as an expansion in powers of $1/N$~\cite{tHooft:1973alw}. 

The small coupling on its own is not enough to guarantee a small production rate.  Typically, cross sections scale with the 't Hooft coupling which means that any individual final state is $N$-suppressed, but the sum over all states can result in an $\cO(1)$ rate.  One exception is when only a subset of the new particles, $N_{\rm LHC}$, are kinematically-accessible. In this case total rates can be suppressed by powers of $N_{\rm LHC}/N$, as illustrated in Fig.~\ref{fig:scheme1}.
 
This leaves us with the last and most interesting aspect of these new sectors: long decay chains.  If the new sector is somewhat secluded, {\it i.e.} the couplings to the SM are smaller than the couplings between the new states, it is likely for states towards the top of the spectrum to cascade within their own sector before decaying to the SM.  In a dense spectrum there is no phase space suppression for not decaying directly to the bottom of the spectrum. 

Furthermore, in dense spectra the production rate of states higher in the spectrum can be comparable to that of the lowest-lying states.  Since the only guaranteed low multiplicity final states come from the production of the lightest particle, in such spectra the probability is $\cO(1/N_{\rm LHC})$ for single production and $\cO(1/N_{\rm LHC}^2)$ for pair production so that higher multiplicity final states can dominate the total production cross section.

%%%%%%%%%%%%%%%%%%%%%%%%%%%%%%%%%%%%%%%
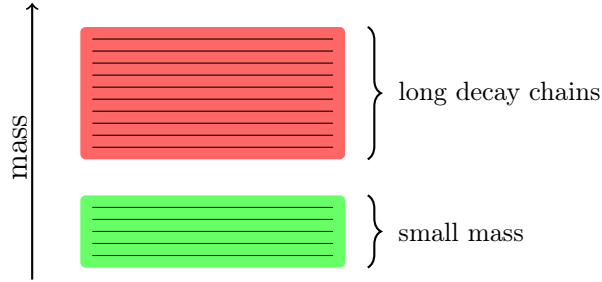
\begin{figure}[!t]
\centering
  \begin{tikzpicture}[scale=1.6]
  \draw[->,line width=0.8pt] (-0.5 ,-1.1) -- (-0.5 , 1.2);
  \node[rotate=90] at (-0.6, 0.0 ) {mass};
  %%%
  \filldraw[draw=none,fill=red, fill opacity=0.6, rounded corners=2pt]   (-0.1 ,-0.1 ) rectangle ( 2.1 , 1.0 ) ;
  \filldraw[draw=none,fill=green, fill opacity=0.6, rounded corners=2pt] (-0.1 ,-0.4 ) rectangle ( 2.1 ,-1.0 ) ;
  \draw (0,0.9) -- (2.0,0.9);
  \draw (0,0.8) -- (2.0,0.8);
  \draw (0,0.7) -- (2.0,0.7);
  \draw (0,0.6) -- (2.0,0.6);
  \draw (0,0.5) -- (2.0,0.5);
  \draw (0,0.4) -- (2.0,0.4);
  \draw (0,0.3) -- (2.0,0.3);
  \draw (0,0.2) -- (2.0,0.2);
  \draw (0,0.1) -- (2.0,0.1);
  \draw (0,0.0) -- (2.0,0.0);
  \draw (0,-0.5) -- (2.0,-0.5);
  \draw (0,-0.6) -- (2.0,-0.6);
  \draw (0,-0.7) -- (2.0,-0.7);
  \draw (0,-0.8) -- (2.0,-0.8);
  \draw (0,-0.9) -- (2.0,-0.9);
  %%%%
  \draw [decorate,decoration={brace,amplitude=5pt,mirror,raise=4pt},yshift=0pt,line width=0.8pt] (2.2,-1.0) -- (2.2,-0.4) 
        node [black,midway,xshift=0.4cm,right] {\footnotesize small mass};
  \draw [decorate,decoration={brace,amplitude=5pt,mirror,raise=4pt},yshift=0pt,line width=0.8pt] (2.2,-0.1) -- (2.2, 1.0) 
        node [black,midway,xshift=0.4cm,right] {\footnotesize long decay chains};
  \end{tikzpicture}
\captionsetup{width=0.87\linewidth}
\caption{A theory with a moderately large number of new particles in a finite mass range, having a small coupling to the SM, is characterized by a few light states that decay directly to the SM and events with high multiplicities produced by the long decay chains of the heavier states.}
\label{fig:scheme2}
\end{figure}
%%%%%%%%%%%%%%%%%%%%%%%%%%%%%%%%%%%%%%%

Observing states near the bottom or top of the spectrum each have their own challenges.  If we produce a state towards the bottom of the spectrum, most of the time it decays directly to the SM, generating a low-multiplicity final state already targeted by current searches.  However these particles might just be too light to be detected either because of trigger thresholds or because of backgrounds, especially if they decay to jets. Furthermore, the total rate for these low multiplicity final states might just be too small to be detectable. 

On the contrary a particle near the top of the spectrum cascades within its sector before decaying back to a large number of SM particles.  The total visible and/or invisible energy can easily be too small to fire an $H_T$ or missing energy trigger if all the new particles are relatively light ({\it i.e.} a few hundreds of GeV rather than a few TeV) and/or the spectrum is compressed. This is summarized schematically in Fig.~\ref{fig:scheme2}.

In the next section we introduce a model that makes this simple discussion more concrete and in Sect.~\ref{sec:pheno} we study the statements made in the previous paragraphs quantitatively.  We consider sectors with masses around a few hundreds of GeV with $N$ between 5 and 50, but obviously the statements made here are much more general.

%**************** Section ***********************************
\section{A Concrete Model}
\label{sec:models}
%*************************************************************

In this section we introduce an explicit model.  Other models can be constructed, for example the one in Sect.~\ref{sec:fermions}, but this one is the simplest and illustrates all relevant features. Consider $N$ real scalars with the Lagrangian
\begin{equation} \label{eq:scalars}
\mathcal{L}_{\phi} = \frac{1}{2}\partial_\mu \phi_i \partial^\mu \phi_i-\frac{m_i^2}{2} \phi_i^2 - a_{i j k}\phi_i \phi_j \phi_k- \lambda_{i j k l} \phi_i \phi_j \phi_k \phi_l ,
\end{equation}
where sums over repeated indices are implied and run from $1$ to $N$.  The parametrization should be interpreted as an expansion around a local minimum, not necessarily valid for arbitrarily large field excursions.

In addition to Eq.~(\ref{eq:scalars}), we connect the new scalars to the SM through the most relevant interactions that are possible, the Higgs portal couplings,
\begin{equation} \label{eq:higgscoupling}
\mathcal{L}_{\phi H} = -a^{\rm SM}_i \phi_i |H|^2 - \lambda^{\rm SM}_{ij} \phi_i \phi_j |H|^2,
\end{equation}
where again sums over repeated indices are left implicit.  In Sect.~\ref{sec:pheno}, where we discuss collider phenomenology, we consider models with $N$ ranging from 5 to 50.
We study separately three distinct phenomenological possibilities:
\begin{enumerate}
\item $a_{ijk}=a^{\rm SM}_i=0$.
\item All interactions are present and single production dominates.
\item All interactions are present and pair production dominates.
\end{enumerate}
The first case has a potential that respects a $Z_2$ symmetry under which
\begin{equation}
\phi_i \to -\phi_i,
\quad\quad\quad
\forall i.
\end{equation}
It can be considered our ``nightmare scenario'' being maximally difficult to detect at colliders. The second one is a more faithful description of a friendly landscape~\cite{ArkaniHamed:2005yv}, where the new scalars can get vacuum expectation values of the order of their masses. The third possibility covers a different limit of this model that has distinct phenomenological features.  From the point of view of the Lagrangian it is similar to the first case, but with the addition of a small $Z_2$-breaking.  The breaking is sufficiently small that pair production is still the dominant production process at the LHC.

The model defined by Eqs.~\eqref{eq:scalars} and~\eqref{eq:higgscoupling} maps onto models used for baryogenesis (with a particular choice of interactions)~\cite{Meade:2018saz, Baldes:2018nel, Glioti:2018roy} and can be a QFT model of the landscape~\cite{Bardeen:1985tr,ArkaniHamed:2005yv,Bray:2007tf,Easther:2016ire,Dine:2015ioa}. The new scalars can also be moduli from extra dimensions compactified at some large scale $M_*$.  If the compactification scale is much larger than their mass, they would have to be tuned moduli to be visible at colliders, {\it i.e.} they need couplings not suppressed by some power of $m_\phi/M_*$.  Given the ubiquitous presence of the weak scale in nature their presence around LHC energies might not be a coincidence.  The dark matter energy density today is $\rho_{\rm DM}\sim (v^2/M_{\rm Pl})^4$, the cosmological constant and neutrino masses are also related to the same combination of $v$ and $M_{\rm Pl}$: $\Lambda_{\rm CC}^{1/4} \sim m_\nu \sim v^2/M_{\rm Pl}$, not to mention the role of the weak scale in the SM itself. From a more pragmatic perspective, this might be just one of many sectors spread over many orders of magnitude in mass that arise from the compactification of extra dimensions and supersymmetry breaking.

%*************************************************************
\subsubsection*{Large $N$} 

To have a well-behaved perturbative theory when $N \gg 1$ the couplings must scale as an inverse power of $N$.  First considering only the quartic interactions, we need that $\lambda^{\rm SM} \lesssim 4\pi/N$ and $\lambda \lesssim 16\pi^2/N^2$.  This can be verified by inspecting diagrams.  For the kind of arguments needed to derive this scaling in general see Refs.~\cite{Coleman:1985rnk,Cohen:2018cnq}.

When trilinear couplings are included we require that $a^{\rm SM} \lesssim 4 \pi v / \sqrt{N}$ and $a \lesssim 4 \pi m_\phi / N^{3/2}$, where $m_\phi$ is the typical mass scale of the scalars and $v\approx 174$~GeV is the vacuum expectation value of the SM Higgs.  

%*************************************************************
\subsubsection*{Randomness} 

We introduce phenomenological randomness into the model through the mass matrix.\footnote{Randomness in the mass matrix was also studied in~\cite{Dienes:2016kgc}.}  We use fully-populated matrices where each entry is drawn randomly from a Gaussian distribution.  For $N$ real scalars the mass matrix is then symmetrized.  After diagonalization such a random matrix has a known distribution for the mass eigenvalues in the large $N$ limit.  In Eq.~\eqref{eq:scalars} we write the model after diagonalization.

We take the trilinear and the quartic couplings to be roughly of the same order.  This reflects our choice of a fully-populated mass matrix with entries drawn from the same distribution, as in this case all mass eigenstates overlap at $\cO(1/\sqrt{N})$ with all flavor eigenstates.

%*************************************************************
\subsubsection*{Existing Constraints} 

The most stringent bound on these models is indirect and arises from LHC measurements of Higgs couplings. The mixing between the Higgs and the new scalars reduces all Higgs couplings, modifying the global signal strength measured at the LHC.  We can estimate the impact of this bound by taking all scalar masses to be approximately $m_\phi$.  The modification of the couplings of the Higgs is governed by
\begin{equation} \label{eq:mixing}
\sin\theta \approx \frac{a^{\rm SM} v \sqrt{N}}{m_\phi^2-m_h^2}.
\end{equation}
A weighted average of the CMS and ATLAS global signal strengths from 7, 8, and 13 TeV data~\cite{Khachatryan:2016vau, ATLAS:2018doi, Sirunyan:2018koj} sets a limit of $\sin\theta < 0.13$. The values of $a^{\rm SM}$ that we consider in Sect.~\ref{sec:pheno} are all largely consistent with this bound.

Electroweak precision measurements receive corrections at one-loop~\cite{Barbieri:2007bh}
\begin{equation} \label{eq:ewpt}
\hat S \approx \frac{\alpha}{48 \pi s_w^2}\sin^2\theta \log \frac{m^2_\phi}{m_h^2},
\quad\quad\quad 
\hat T \approx -\frac{3\alpha}{16\pi c_w^2}\sin^2\theta \log \frac{m^2_\phi}{m_h^2},
\end{equation}
and are constrained at the permille level making them subdominant~\cite{Buttazzo:2015bka}.  Above we have again imagined all scalars to have the same mass $m_\phi$.

The new states can also be singly produced at colliders through the mixing with the Higgs.  The most relevant direct searches are those for heavy Higgses, but existing searches, once the bound from Eq.~\eqref{eq:mixing} is taken into account, have no residual exclusion power~\cite{Aaboud:2017yyg, Aaboud:2017gsl, Aaboud:2017rel, Khachatryan:2015cwa, CMS:2015mca}.  However, the full 13 TeV dataset has not yet been analyzed and searches for resonant pairs of electroweak gauge bosons will become relevant in the future.

In the $Z_2$-symmetric case the new scalars do not mix with the Higgs boson and the constraints that we have just discussed do not apply. Furthermore, the new particles can only be pair produced at colliders with much smaller rates.  Current searches for pairs of new particles have not even begun to probe $\lambda^{\rm SM} \sim \mathcal{O}(1)$~\cite{Aaboud:2017nmi, Sirunyan:2018rlj,Sirunyan:2018pwn, CMS:2012iya}. 

%**************** Section ***********************************
\section{Disordered Model Building}
\label{sec:disorder}
%*************************************************************

In this section we discuss the spectrum and eigenvectors of disordered mass matrices.  We focus on $N \times N$ matrices with entries randomly drawn from Gaussian distributions.  
Our starting point is the Gaussian Orthogonal Ensemble (GOE), defined by
\begin{equation} \label{eq:jpdfsimple}
\rho[M]=
\prod_{i=1}^N \frac{e^{-(M_{ii})^2/2}}{\sqrt{2\pi}}
\prod_{i<j} \frac{e^{-(M_{ij})^2}}{\sqrt{\pi}},
\end{equation}
where we use $M_{ii}$ to denote the diagonal entries of the mass matrix $M$ and $M_{ij}$ for the off-diagonal entries.  In Eq.~\eqref{eq:jpdfsimple} the entries are drawn from the standard Gaussian distribution $\cN(0,1)$ with mean $\mu = 0$ and standard deviation $\sigma = 1$.\footnote{The standard deviation of the off-diagonal entries is $1/\sqrt{2}$ rather than $1$ because we imagine that the matrix is symmetrized via $(M+M^T)/2$.} The GOE contains real symmetric matrices which correspond to models of real scalars.  For complex scalars the analogous ensemble is the Gaussian Unitary Ensemble (GUE) that contains complex Hermitian matrices.  Most of our results apply to both cases (where there are differences we specify them explicitly).

The natural generalization of the GOE, with broader physical applications, that we consider is
\begin{equation} \label{eq:jpdf}
\rho[M]=
\prod_{i=1}^N \frac{e^{-\frac{(M_{ii}-\mu_d)^2}{2 \sigma_d^2}}}{\sqrt{2\pi}\sigma_d}
\prod_{i<j} \frac{e^{-\frac{(M_{ij}-\mu_o)^2}{2\sigma_o^2}}}{\sqrt{2\pi}\sigma_o} . 
\end{equation}
where the diagonal entries are now drawn from $\cN(\mu_d,\sigma_d)$ and the off-diagonal entries from $\cN(\mu_o,\sigma_o)$.  When we refer to the parameters as $\mu$ and $\sigma$ we are setting $\mu = \mu_d = \mu_o$ and $\sigma = \sigma_d = \sigma_o$ and drawing all entries from $\cN(\mu,\sigma)$.

%*************************************************************
\subsection{Summary of Main Results}
\label{sec:dis_summary}

The main result of this section is that in the large $N$ limit the spectral density of matrices with Gaussian-random entries follows a universal distribution called the Wigner semicircle distribution~\cite{WignerI, WignerII, Dyson:1962es}, which once appropriately normalized reads
\begin{equation} \label{eq:wigner}
\rho_{\rm SC}(m) = \frac{1}{2\pi}\sqrt{4-m^2}.
\end{equation}
For finite $N$ and including the dependence on the parameters used for the Gaussian distribution the spectral density is
\begin{equation} \label{eq:wignerWithParameters}
\rho(m) = \frac{1}{\pi \beta N \sigma^2}\sqrt{2 \beta N \sigma^2 - m^2} + \cO(1/N).
\end{equation}
The parameter $\beta$ is called the Dyson index and specifies the ensemble used.  For the GOE we have $\beta=1$ and for the GUE we have $\beta=2$.  The parameter $\mu$ does not appear in Eq.~\eqref{eq:wignerWithParameters} as the distribution does not depend on $\mu$ with the exception of a single large eigenvalue at $\sim N \mu$.  The spectral edges are at $\pm \sqrt{2 \beta N} \sigma$.

\begin{figure}[!t]
\begin{center}
\includegraphics[width=0.4\textwidth]{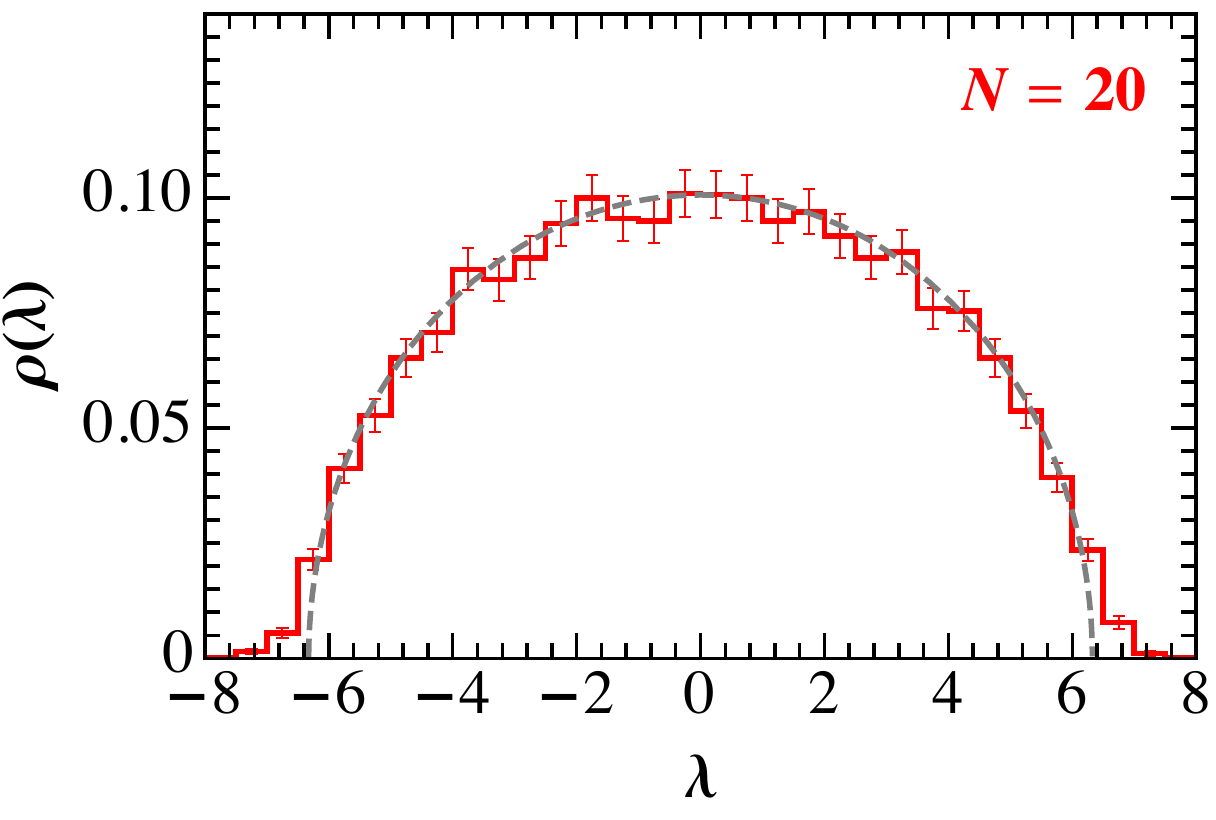} \quad\quad\quad
\includegraphics[width=0.4\textwidth]{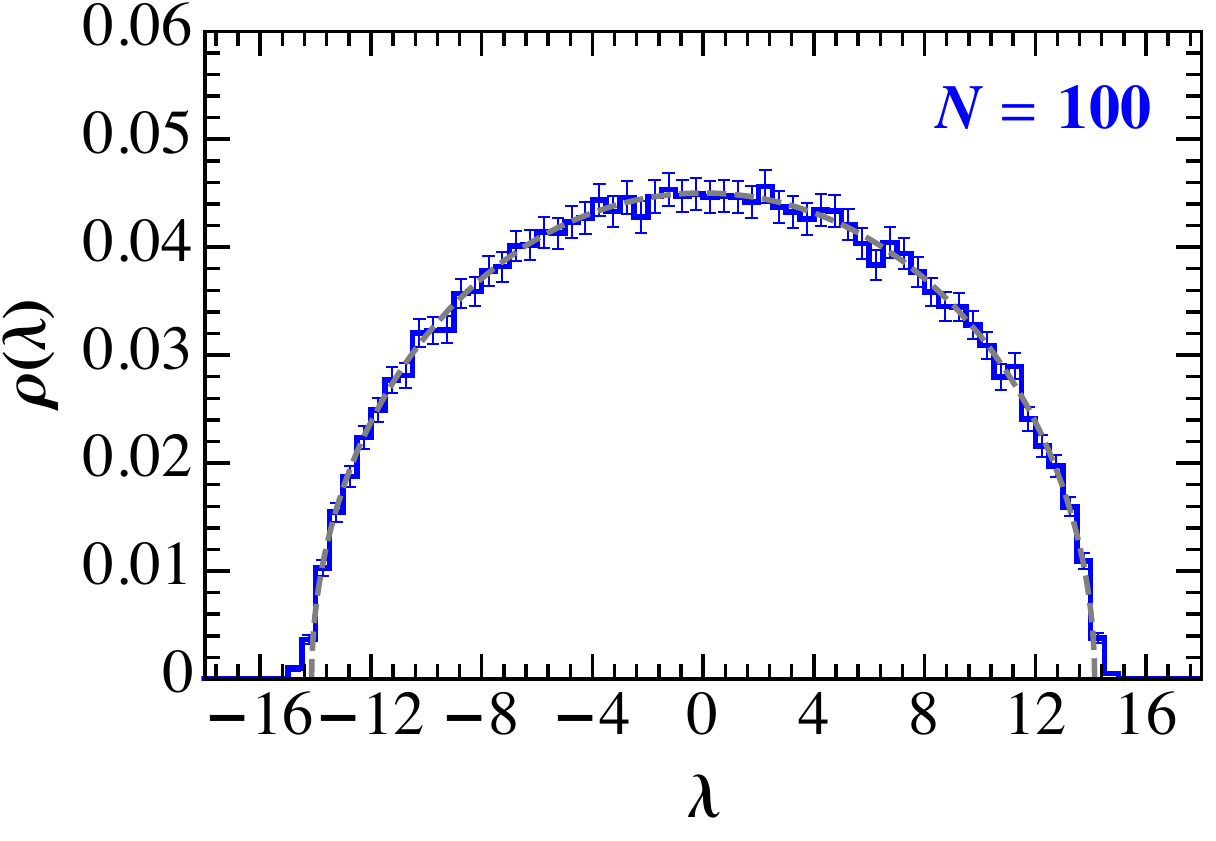}
\captionsetup{width=0.87\linewidth}
 \caption{Left: the spectral density of a $20 \times 20$ matrix (red) drawn from the Gaussian Orthogonal Ensemble and the Wigner semicircle (gray).  Right: the spectral density of a $100 \times 100$ matrix (blue) drawn from the Gaussian Orthogonal Ensemble and the Wigner semicircle (gray).}
 \label{fig:Wigner}
 \end{center}
\end{figure}

In Fig.~\ref{fig:Wigner} we plot the eigenvalues of matrices with $N=20$ and $N=100$ along with the Wigner semicircle distribution.

The eigenvectors of these matrices do not significantly deviate from what one might naively expect.  Since the mass matrix is fully-populated with elements of comparable magnitude, we can imagine the eigenvectors to be an approximately democratic mix of flavor eigenstates with weight $\sim 1/\sqrt{N}$.  This is indeed the case as the sum of the squares of the components of the GOE eigenvectors follow a beta distribution with mean $\sim 1/N$ and variance $\sim 1/N^2$~\cite{ORourke2016a}.

We can infer several consequences from the Wigner semicircle distribution:
\begin{itemize}
\item All eigenvalues (except the largest one) are roughly contained in the interval $(- \sqrt{N} \sigma, \sqrt{N} \sigma )$.  In particular if $\sigma \ll \mu$ they can be much smaller than the scale $\mu$ of the matrix entries.
\item The typical eigenvalue spacing is $\sim \sigma/\sqrt{N}$.  
\item The lightest eigenvalue is $\sim \sigma/\sqrt{N}$.
\end{itemize}

%%%%%%%%%%%%%%%%%%%%%%%%%%%%%%%%%%%%%%%%
%%%%%%%%%%%%%%%%%%%%%%%%%%%%%%%%%%%%%%%%

The fact that $N-1$ of the eigenvalues fall in $(- \sqrt{N} \sigma, \sqrt{N} \sigma )$ even when $\mu \gg \sigma$ appears useful for model building.  If one found a model where $\mu \sim M_{\rm Pl}$ while $\sigma \sim v$ it would appear that a hierarchy has been generated.  However we have only rewritten $N-1$ Goldstone bosons in an unusual basis.

To see this, consider $N$ real scalars contained in a vector $\Phi$ with an $O(N)$-symmetric potential
\begin{equation}
V(\Phi) = - \frac{m^2}{2} \Phi^T \Phi + \frac{\lambda}{4} (\Phi^T \Phi)^2 .
\end{equation}
If we expand $V(\Phi)$ around the true minimum and take the VEV in the direction $\langle \Phi \rangle = (v, \ldots,  v)^T$ the mass matrix of the physical degrees of freedom is precisely what we obtain in our examples by sending the standard deviation $\sigma$ to zero, {\it i.e.} a matrix with all equal entries.  This matrix has $N-1$ massless eigenvalues which are the Goldstone bosons of SO($N$)/SO($N-1$). The large eigenvalue with mass squared proportional to $N$ is the radial mode.  This explains why the typical scale of the eigenvalues is $\sigma$ which is explicitly breaking the symmetry, instead of $\mu$ that preserves it.

The last useful result on the spectrum is that taking a different mean for the diagonal elements $\mu_d$, shifts the spectral density from zero by the amount $(\mu_d-\mu)$
\begin{equation}
\rho(m) = \frac{1}{\pi \beta N \sigma_o^2}\sqrt{2 \beta N \sigma_o^2 - (m-(\mu_d-\mu))^2} + \cO(1/N) .
\end{equation}
If viewed in terms of Goldstone bosons this is a consequence of the explicit breaking of the $O(N)$ symmetry. 

Before turning to the derivation of these results it is worth commenting on a difference between our models and disordered condensed matter systems. It is natural to ask if we can pick any variance for our mass matrix or if we need something scaling as an inverse power of $N$ to have a consistent theory. In real disordered systems locality and the central limit theorem imply that the free-energy is self-averaging, {\it i.e.}
\begin{equation}
\langle F^2 \rangle - \langle F \rangle^2 = \mathcal{O}(1/N),
\end{equation}
where $\langle \cdot \rangle$ is a disorder average. The argument is quite intuitive: only particles that are nearby interact with each other appreciably and we can divide $F$ into a sum over many cells that are not strongly interacting with each other. In the large $N$ limit we can ignore surface effects and recover the result above. 

If this result applied to us, it would require for example $\sigma^2 \sim 1/N$ for our scalars, just from computing the free energy from the partition function $Z$
\begin{equation}
F = - T \log Z .
\end{equation}
However, in our theories there is no notion of locality in the same sense as for condensed matter systems. The scalar with flavor $1$ can mix as strongly with that of flavor $2$ as with any other, so in this sense all interactions can be long range.

%*************************************************************
\subsubsection*{Catalan Numbers}

There is another interesting structural property of our large $N$ sectors with Gaussian-random mass matrices.  The set of numbers
\begin{equation}
C_n = \frac{1}{n+1} \binom{2n}{n} ,
\end{equation}
known as the Catalan numbers, impacts both the shape of the eigenvalue distribution and the length of our decay chains.  The moments of the Wigner semicircle distribution are the Catalan numbers
\begin{equation}
C_n = \int_{-2}^2 \rho_{\rm SC}(m)m^{2n} dm .
\end{equation}
A derivation of this result can be found in App.~\ref{app:moments} while an explanation that uses an interesting connection with planar diagrams is presented in App.~\ref{app:catalan}. 

To see the role of the Catalan numbers for our decay chains,  consider a scalar sector with particles that can either decay to two other scalars or to two SM particles. This is a good approximation of our general model in Sect.~\ref{sec:models} since we expect the trilinear couplings to dominate the branching ratios.

After a first decay, each daughter scalar would then likewise either decay to additional scalars or to two SM particles.  Each possible final state can then be represented by a binary tree.  The average number of final state particles can be approximated by computing the weighted sum of possible final states.  This requires knowing, for a given number of leaves $2n$, how many distinct binary trees there are.  This sequence of numbers is again given by $C_n$.

The asymptotic behavior of the Catalan numbers also lets us make a rudimentary estimate of the average decay chain length.  Let the probability of decaying to the SM be $p$ and the probability of decaying to the new sector be $q$.  The average number of final state particles, ignoring all phase space factors, is then 
\begin{equation}
\sum_{n=0}^{N_{\rm max}} (n+1) C_n (pq)^n p,
\end{equation}
with $N_{\rm max}$ determined by phase space. For $n \to \infty$ we have that $C_n \sim 4^n / n^{3/2}$ which means that the $n$th term in the average goes like $\sim (4pq)^n /\sqrt{n}$.  Assuming that $(4pq) \sim \cO(1)$ the average will go like $\sim \sqrt{N_{\rm max}}$.  $N_{\rm max}$ is set by phase space and is $\cO(100)$ for the parameters considered in Sect.~\ref{sec:pheno} in good agreement with the numerical results in the same section. This heuristic derivation is valid for two-body decays, but the scaling of $\sim \sqrt{N_{\rm max}}$ continues to apply for higher $n$-body decays.

%*************************************************************
\subsection{The Joint Eigenvalue Distribution}
\label{sec:jointeigenvalue}

Most of the results in the previous section can be derived using path integral techniques and can be estimated by simple dimensional analysis. In this section we go through the heuristic arguments that justify the form of the spectrum.

The first step is to go from the joint entry distribution $\rho[M]$, in Eq.~\eqref{eq:jpdf}, to the joint eigenvalue distribution $\hat{\rho}[m_1, \ldots, m_N]$ via the change of basis
\begin{equation}
M = U M_D U^\dagger,
\end{equation}
where $M_D={\rm diag}(m_1, \ldots, m_N)$.  The joint distribution becomes
\begin{equation} \label{eq:pdftransform}
\rho[M] DM = \rho[M_D] |J(M_D)| Dm DU.
\end{equation}
Note that $\rho[M_D]$ is not the joint eigenvalue distribution because the Jacobian $|J(M_D)|$ remains.  The joint eigenvalue distribution is rather $\hat{\rho}[m_1, ..., m_N] = \rho[M_D]|J(M_D)| \int dU$.

The metric on the space of symmetric (or Hermitian) matrices that defines $DM$ is induced by the product $M_1 \cdot M_2 = {\rm Tr}[M_1 M_2]$, {\it i.e.}
\begin{equation}
ds^2_M = {\rm Tr}[dM dM],
\end{equation}
where $ds^2_M$ is the distance on the space of matrices. The other differentials are
\begin{equation}
Dm = \prod_i dm_i, 
\quad\quad\quad
DU=\prod_{i>j} (U^\dagger dU)_{ij} .
\end{equation}
For simplicity we start with the case $\mu = 0$.  Eq.~\eqref{eq:pdftransform} anticipates that the Jacobian of the transformation $|J(M_D)|$ and $\rho[M_D]$ depend only on the eigenvalues.  The proofs are given below and in App.~\ref{app:vandermonde}. If we think about $\rho$ as an action for a matrix model, the eigenvectors $U$ represent the gauge freedom associated to the choice of basis and $J$ is a gauge-invariant Fadeev-Popov determinant. Then it is not surprising that $J$ does not depend on $U$.

If $J$ depends only on the eigenvalues we can make an ansatz for its form based on the following arguments. When two eigenvalues coincide the transformation becomes singular and $J$ must be zero, so we expect
\begin{equation}
|J|=\prod_{i<j}|m_i-m_j|^\beta .
\end{equation}
To determine $\beta$ we can use dimensional analysis\footnote{We use square brackets $[\cdot]$ to indicate the dimensions of a quantity.}
\begin{equation}\begin{aligned}
{\rm GOE}: & \quad [DM]=m^{N(N+1)/2},   & [|J(M_D)| Dm]=m^N m^{\beta N(N-1)/2}, \\
{\rm GUE}: & \quad [DM]=m^{N^2},        & [|J(M_D)| Dm]=m^N m^{\beta N(N-1)/2},   
\end{aligned}\end{equation}
from which we find that $\beta=1$ for the GOE and $\beta=2$ for the GUE.  Now we are left with evaluating $\rho[M_D]$. If we take $\mu = 0$ in Eq.~\eqref{eq:jpdf} it is easy to conclude that in the large $N$ limit
\begin{equation}
 \rho[M] \propto e^{-\frac{1}{2\sigma^2}{\rm Tr}[M^2]+\mathcal{O}(1/N)} = e^{-\frac{1}{2\sigma^2}\sum_{i=1}^N m_i^2+\mathcal{O}(1/N)}\, .
\end{equation}
Combining this with the result for the Jacobian we finally obtain the joint distribution of eigenvalues
\begin{equation}\label{eq:joint}
\hat{\rho}[m_1, \ldots,  m_N)] 
\equiv \rho(M_D)|J(M_D)|
= \frac{1}{Z_{N, \beta}} e^{-\frac{1}{2\sigma_o^2}\sum_{i=1}^N m_i^2} \prod_{i<j}|m_i-m_j|^\beta,
\end{equation}
where $Z_{N, \beta}$ is a normalization factor.

From Eq.~\eqref{eq:joint} is possible to derive the Wigner semicircle distribution, Eq.~\eqref{eq:wigner}, by solving a path integral~\cite{2017arXiv171207903L, Zee:2003mt}.  We can either use Feynman diagrams, as done in App.~\ref{app:catalan}, or use a saddle point approximation~\cite{WignerI, Dyson:1962es}. Even without going through the derivation we can understand most of the results in the previous section just by looking at Eq.~\eqref{eq:joint}.

For example we expect the largest positive and negative eigenvalues to be $\cO(\pm\sqrt{\beta N}\sigma)$ just from expanding $\hat{\rho}[m_1, \ldots, m_N]$ around the largest eigenvalue $m_*$ 
\begin{equation}
\hat{\rho}[m_1, \ldots, m_N] \sim e^{-\frac{1}{2\sigma^2} m_*^2} |m_*|^{\beta (N-1)} ,
\end{equation}
and taking the derivative $d\hat{\rho}/dm_*$ to find the maximum.

Thus far we have assumed $\mu = 0$ but it can be shown that the results are valid also for $\mu \neq 0$.  Consider splitting $M$ into a zero-mean matrix $M^\prime$ and a constant matrix $A$
\begin{equation}
M = M+A-A = M^\prime + A.
\end{equation}
If we apply a unitary transformation $U_A$ that diagonalizes $A$ we find that $U_A A U_A^\dagger = {\rm diag}(0, \ldots, 0, N\mu)$.  At the same time $M^\prime$ is rotated to $M'' = U_A M^\prime U_A^\dagger$.  The asymptotic distribution of eigenvalues of $M^{\prime\prime}$ is still the Wigner semicircle distribution and even the finite $N$ joint distribution is the same.  The reason is that $\rho[M]$ is invariant under unitary transformations of $M$ and the Jacobian is the same for $M$ and $M^{\prime\prime}$ since they are both Hermitian.\footnote{For a derivation of this result see App.~\ref{app:vandermonde}}

A similar argument can be applied to the case with a different mean for the off-diagonal entries: $\mu_d \neq \mu_o$.   We can subtract the matrix $B$ where $B$ contains the mean of each entry.  Then we further split $B$ into a constant matrix $A$ and $(\mu_d-\mu_o)$ times the identity
\begin{equation}
M= M+B-B=M^\prime+B= M^\prime+A+(\mu_d-\mu_o)\mathbf{1}_{N\times N}.
\end{equation}
By diagonalizing $A$, $M^\prime$ is rotated as before, while the term proportional to the identity, $(\mu_d-\mu_o)\mathbf{1}_{N\times N}$, is unaffected.  Therefore we still have the Wigner semicircle distribution, but shifted from zero by an amount $(\mu_d-\mu_o)$.  In addition we have the large eigenvalue which is also shifted to $m_{\rm max}= N(\mu_o-1) + \mu_d$. 

%%%%%%%%%%%%%%%%%%%%%%%%%%%%%%%%%%%%%%%%
\begin{figure}[!t]
\centering
\includegraphics[width=0.6\textwidth]{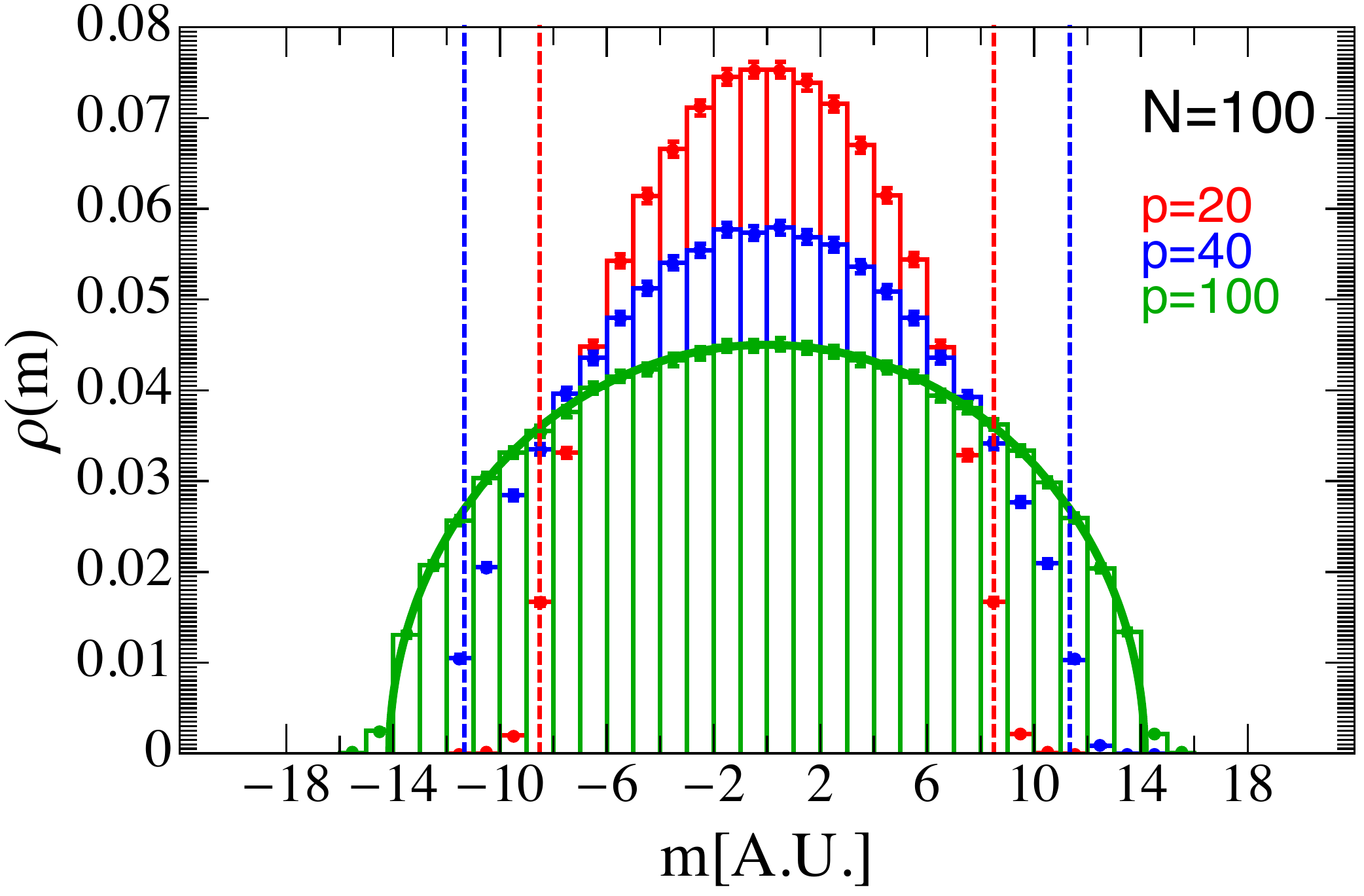}
\captionsetup{width=0.87\linewidth}
 \caption{Spectral density of $100 \times 100$ symmetric band matrices with elements drawn randomly from Gaussian distributions. We have used $\mu_d=\mu_o=0$ and $\sigma_d=\sqrt{2}\sigma_o=1$.  The different colors correspond to different numbers of interacting neighbors $p$. The vertical lines show the estimate of the distribution width given in Eq.~\eqref{eq:betap} from dimensional analysis.}
\label{fig:bandeigenvalues}
\end{figure}
%%%%%%%%%%%%%%%%%%%%%%%%%%%%%%%%%%%%%%%%

This concludes our heuristic derivation of results on the spectrum of GOE and GUE matrices. We now turn to another phenomenological possibility: having mass matrices that mix only $p$ nearest neighbors.

%*************************************************************
\subsection{Band Matrices}

It is interesting to consider what happens if we draw the elements of our matrices from the same probability distributions considered in the previous sections, but instead of populating the full matrix we allow only for nearest neighbor interactions among $p<N$ nearest neighbors
\begin{equation} \label{eq:bandmatrix}
M_p^2 = \left(\begin{array}{cccccc}
m_{11}^2 & m_{12}^2 & \cdots & m_{1p}^2 & \cdots & 0            \\
m_{21}^2 & m_{22}^2 & \cdots & m_{2p}^2 & \cdots & 0            \\
\vdots   & \vdots   & \ddots & \cdots   & \cdots & \vdots       \\
m_{p1}^2 & m_{p2}^2 & \vdots & \ddots   & \cdots & \vdots       \\
\vdots   & \vdots   & \vdots & \vdots   & \ddots & \vdots       \\
0        & 0        & \cdots & \cdots   & \cdots & m_{NN}^2
\end{array}\right).
\end{equation}
This scenario can arise, for example, from localization in an extra dimension.  The eigenvalues of these matrices are spread over a smaller range than those in the GOE or GUE.  $\rho[M]$ is the same as before leaving intact the Gaussian measure in Eq.~\eqref{eq:joint}, but the Jacobian of the trasformation, even just on dimensional grounds, cannot be the same.  If we repeat the arguments in the previous subsection we find
\begin{equation}\begin{aligned}
{\rm GOE}_p: & \quad [DM]=m^{(2N-p)(1+p)/2},   & \quad [|J(M_D)| Dm]=m^N m^{\beta_p N(N-1)/2}.
\end{aligned}\end{equation}
From this we find that
\begin{equation} \label{eq:betap}
\beta_p =\frac{p(p+1-2N)}{N(1-N)}\approx \frac{2 p}{N}.
\end{equation}
In this case the eigenvalues are spread over an interval that we expect to be $\cO(\sqrt{p}\sigma_o)$ wide. This is indeed confirmed numerically, as shown in Fig.~\ref{fig:bandeigenvalues}.

Also the eigenvectors of these matrices, not surprisingly, are more localized than their GOE or GUE counterparts. The limiting case is a diagonal matrix that exhibits perfect localization. In the opposite limit we might have eigenvectors as spread as in the previous cases. If the off-diagonal elements are smaller than the diagonal ones we can have a stronger form of localization, known as Anderson localization~\cite{Anderson:1958vr}, with mass eigenstates spanning approximately $p$ flavor eigenstates. This fact has already found several applications in the context of BSM physics~\cite{Rothstein:2012hk, Craig:2017ppp} and cosmology~\cite{Green:2014xqa, Brandenberger:2008xc, Zanchin:1997gf}.

We do not explore band matrices in detail in this paper, but from the discussion in this section we expect a collider phenomenology similar to the one that we discuss in Sect.~\ref{sec:pheno} with potentially longer decay chains compared to fully-populated matrices. 

%**************** Section ***********************************
\section{Collider Phenomenology}
\label{sec:pheno}
%*************************************************************

In this section we explore the collider signals predicted by the scalar model that we presented in Eqs.~\eqref{eq:scalars} and~\eqref{eq:higgscoupling}.  In different regions of parameter space quite different behavior is expected ranging from simple dijet resonances (singly or pair produced) to long cascades ending in many SM particles and possibly missing energy. In all cases presented here we extract particle masses from a Wigner semicircle distribution that has support between 100 and 600 GeV.  

%%%%%%%%%%%%%%%%%%%%%%%%%%%%%%%%%%%%%%%%
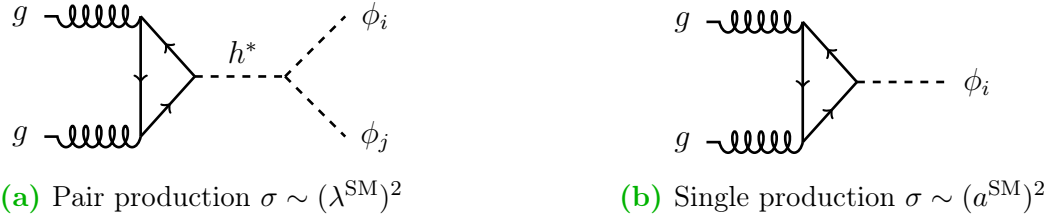
\begin{figure}[!t]\begin{center}
    \begin{subfigure}[t]{0.35\textwidth}\begin{center}
    \begin{tikzpicture}[line width=1.0 pt, scale=0.8]
    \draw[fermion]       ( 0.0, 0.0) -- (-0.9, 1.0);
    \draw[fermion]       (-0.9, 1.0) -- (-0.9,-1.0);
    \draw[fermion]       (-0.9,-1.0) -- ( 0.0, 0.0);
    \draw[gluon]         (-0.9,-1.0) -- (-2.5,-1.0);
    \draw[gluon]         (-0.9, 1.0) -- (-2.5, 1.0);
    \draw[scalarnoarrow] ( 0.0, 0.0) -- ( 1.5, 0.0);
    \draw[scalarnoarrow] ( 1.5, 0.0) -- ( 2.5, 1.0);
    \draw[scalarnoarrow] ( 1.5, 0.0) -- ( 2.5,-1.0);
    \node at (-2.9, 1.0) {$g$};
    \node at (-2.9,-1.0) {$g$};
    \node at ( 3.0, 1.0) {$\phi_i$};
    \node at ( 3.0,-1.0) {$\phi_j$};
    \node at ( 0.8, 0.4) {$h^*$};
    \end{tikzpicture}
    \caption{Pair production $\sigma \sim (\lambda^{\rm SM}) ^2$}
    \label{fig:feynman-prod-pair}
    \end{center}\end{subfigure}
%%%%%%%%%%%%%%%%%%%%%%%%%%%%%%%%%%%%%%%%
    \quad\quad\quad\quad\quad\quad
    \begin{subfigure}[t]{0.35\textwidth}\begin{center}
    \begin{tikzpicture}[line width=1.0 pt, scale=0.8]
    \draw[fermion]       ( 0.0, 0.0) -- (-0.9, 1.0);
    \draw[fermion]       (-0.9, 1.0) -- (-0.9,-1.0);
    \draw[fermion]       (-0.9,-1.0) -- ( 0.0, 0.0);
    \draw[gluon]         (-0.9,-1.0) -- (-2.5,-1.0);
    \draw[gluon]         (-0.9, 1.0) -- (-2.5, 1.0);
    \draw[scalarnoarrow] ( 0.0, 0.0) -- ( 1.5, 0.0);
    \node at (-2.9, 1.0) {$g$};
    \node at (-2.9,-1.0) {$g$};
    \node at ( 2.0, 0.0) {$\phi_i$};
    \end{tikzpicture}
    \caption{Single production $\sigma \sim (a^{\rm SM}) ^2$}
    \label{fig:feynman-prod-single}
    \end{center}\end{subfigure}
%%%%%%%%%%%%%%%%%%%%%%%%%%%%%%%%%%%%%%%%
  \captionsetup{width=0.87\linewidth}
  \caption{Diagrams for scalar production in the model of Eqs.~\eqref{eq:scalars} and~\eqref{eq:higgscoupling}.}
  \label{fig:feynman-production}
\end{center}\end{figure}
%%%%%%%%%%%%%%%%%%%%%%%%%%%%%%%%%%%%%%%%

%%%%%%%%%%%%%%%%%%%%%%%%%%%%%%%%%%%%%%%%
\begin{figure}[h]\begin{center}
    \begin{subfigure}[t]{0.45\textwidth}\begin{center}
    \begin{tikzpicture}[line width=1.0 pt, scale=0.8]
    \draw[scalarnoarrow] ( 0.0, 0.0) -- ( 1.5, 0.0);
    \draw[scalarnoarrow] ( 1.5, 0.0) -- ( 2.5, 1.0);
    \draw[scalarnoarrow] ( 1.5, 0.0) -- ( 3.0, 0.0);
    \draw[scalarnoarrow] ( 1.5, 0.0) -- ( 2.5,-1.0);
    \node at (-0.5, 0.0) {$\phi_i$};
    \node at ( 2.9, 1.2) {$\phi_j$};
    \node at ( 3.4, 0.0) {$\phi_k$};
    \node at ( 2.9,-1.2) {$\phi_l$};
    \end{tikzpicture}
    \caption{Decay via hidden sector quartic $\Gamma \sim \lambda^2$}
    \label{fig:feynman-decay-lambda}
    \end{center}\end{subfigure}
%%%%%%%%%%%%%%%%%%%%%%%%%%%%%%%%%%%%%%%%
    \quad\quad\quad
    \begin{subfigure}[t]{0.45\textwidth}\begin{center}
    \begin{tikzpicture}[line width=1.0 pt, scale=0.8]
    \draw[scalarnoarrow] ( 0.0, 0.0) -- ( 1.5, 0.0);
    \draw[scalarnoarrow] ( 1.5, 0.0) -- ( 2.5, 1.0);
    \draw[scalarnoarrow] ( 1.5, 0.0) -- ( 2.5,-1.0);
    \node at (-0.5, 0.0) {$\phi_i$};
    \node at ( 2.9, 1.2) {$\phi_j$};
    \node at ( 2.9,-1.2) {$h^{(*)}$};
    \end{tikzpicture}
    \caption{Decay via Higgs portal $\Gamma \sim (\lambda^{\rm SM})^2$}
    \label{fig:feynman-decay-lambdaSM}
    \end{center}\end{subfigure} \\
%%%%%%%%%%%%%%%%%%%%%%%%%%%%%%%%%%%%%%%%
    \begin{subfigure}[t]{0.45\textwidth}\begin{center}
    \begin{tikzpicture}[line width=1.0 pt, scale=0.8]
    \draw[scalarnoarrow] ( 0.0, 0.0) -- ( 1.5, 0.0);
    \draw[scalarnoarrow] ( 1.5, 0.0) -- ( 2.5, 1.0);
    \draw[scalarnoarrow] ( 1.5, 0.0) -- ( 2.5,-1.0);
    \node at (-0.5, 0.0) {$\phi_i$};
    \node at ( 2.9, 1.2) {$\phi_j$};
    \node at ( 2.9,-1.2) {$\phi_k$};
    \end{tikzpicture}
    \caption{Decay via hidden sector trilinear $\Gamma \sim a^2$}
    \label{fig:feynman-decay-a}
    \end{center}\end{subfigure}
%%%%%%%%%%%%%%%%%%%%%%%%%%%%%%%%%%%%%%%%
    \quad\quad\quad
    \begin{subfigure}[t]{0.45\textwidth}\begin{center}
    \begin{tikzpicture}[line width=1.0 pt, scale=0.8]
    \draw[scalarnoarrow] ( 0.0, 0.0) -- ( 1.5, 0.0);
    \draw[fermion] ( 1.5, 0.0) -- ( 2.5, 1.0);
    \draw[fermionbar] ( 1.5, 0.0) -- ( 2.5,-1.0);
    \node at (-0.5, 0.0) {$\phi_i$};
    \node at ( 2.9, 1.2) {$b$};
    \node at ( 2.9,-1.2) {$\bar{b}$};
    \end{tikzpicture}
    \caption{Decay via SM trilinear $\Gamma \sim (a^{\rm SM})^2$}
    \label{fig:feynman-decay-aSM}
    \end{center}\end{subfigure}
%%%%%%%%%%%%%%%%%%%%%%%%%%%%%%%%%%%%%%%%
  \captionsetup{width=0.87\linewidth}
  \caption{Diagrams for scalar decays in the model of Eqs.~\eqref{eq:scalars} and~\eqref{eq:higgscoupling}.}
  \label{fig:feynman-decay}
\end{center}\end{figure}
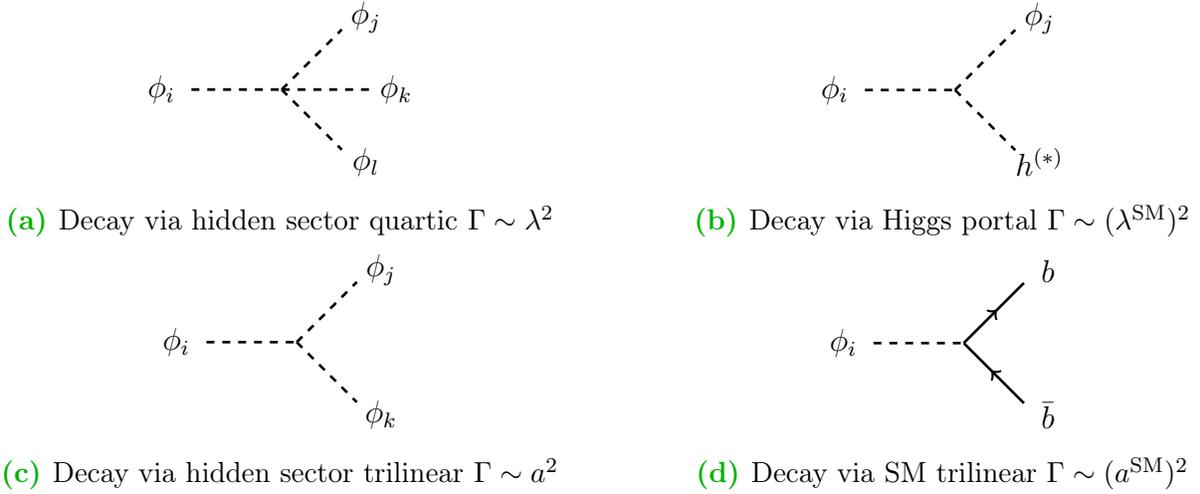
%%%%%%%%%%%%%%%%%%%%%%%%%%%%%%%%%%%%%%%%

We study three regions of parameter space that provide a good representation of the possible final states of the model.  The first case is the $Z_2$-symmetric theory where the only coupling to the SM is a quartic Higgs portal coupling, $\lambda^{\rm SM}$.  The scalars are pair produced through this coupling (Fig.~\ref{fig:feynman-prod-pair}) and can decay either through the same Higgs portal coupling (Fig.~\ref{fig:feynman-decay-lambdaSM}) or through the hidden sector quartic $\lambda$ (Fig.~\ref{fig:feynman-decay-lambda}). For simplicity we take $\lambda^{\rm SM}=1/N$ and $\lambda=1/N^2$ for all the scalars.\footnote{We do not need this scaling to maintain perturbativity in all the cases that we study. We only use it as a convenient benchmark. The results presented here are not strongly affected by this choice of couplings.}

In the second case we allow non-zero trilinear terms to be present. This changes both the production and the dominant decay channels of the scalars.  Production occurs both through pair production (Fig.~\ref{fig:feynman-prod-pair}) and single production (Fig.~\ref{fig:feynman-prod-single}).  The scalars can decay either as before (Fig.~\ref{fig:feynman-decay-lambda} and Fig.~\ref{fig:feynman-decay-lambdaSM}), or through the two-body scalar channel (Fig.~\ref{fig:feynman-decay-a}), or to a pair of SM particles via mixing with the Higgs (Fig.~\ref{fig:feynman-decay-aSM}).  For the quartic couplings we use $\lambda^{\rm SM}=0.1/N$ and $\lambda=1/N^2$. For the trilinears we take $a=m_{\rm min}/N^{3/2}$, where $m_{\rm min}$ is the lightest scalar mass, and $a^{\rm SM}=a$, for all the scalars.

In the last scenario we still allow all the couplings to be present, but we take very small trilinears: $a=10^{-5}\times m_{\rm min}$ and $a^{\rm SM}=10^{-5}\times m_h$, leaving the quartics as in the exactly $Z_2$-symmetric case. In this case pair production dominates over single production and the lightest scalar decays to a pair of SM particles. 

%%%%%%%%%%%%%%%%%%%%%%%%%%%%%%%%%%%%%%%%
\begin{figure}[h]
  \begin{center}
  \includegraphics[width=0.45\textwidth]{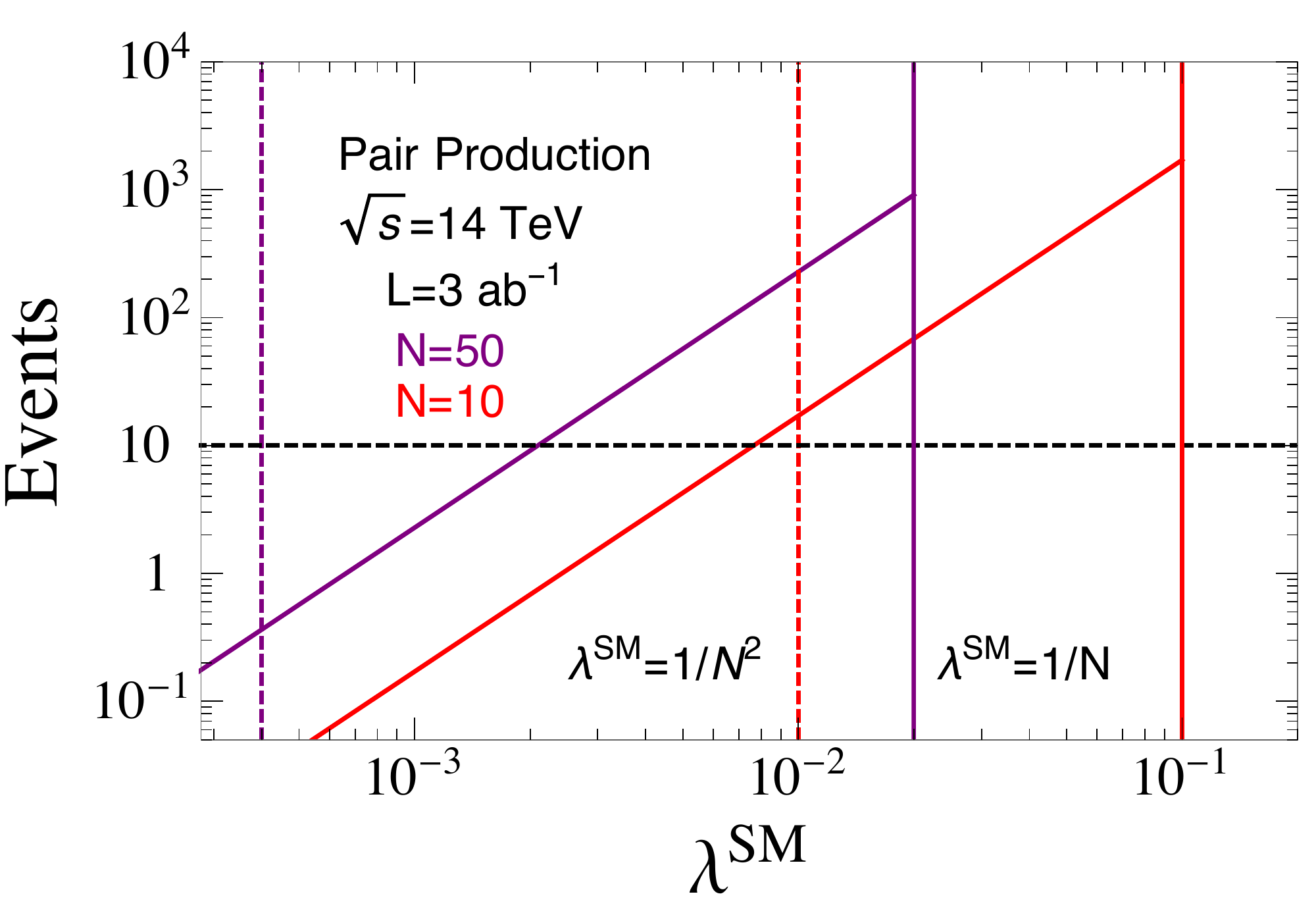} \quad\quad\quad
  \includegraphics[width=0.45\textwidth]{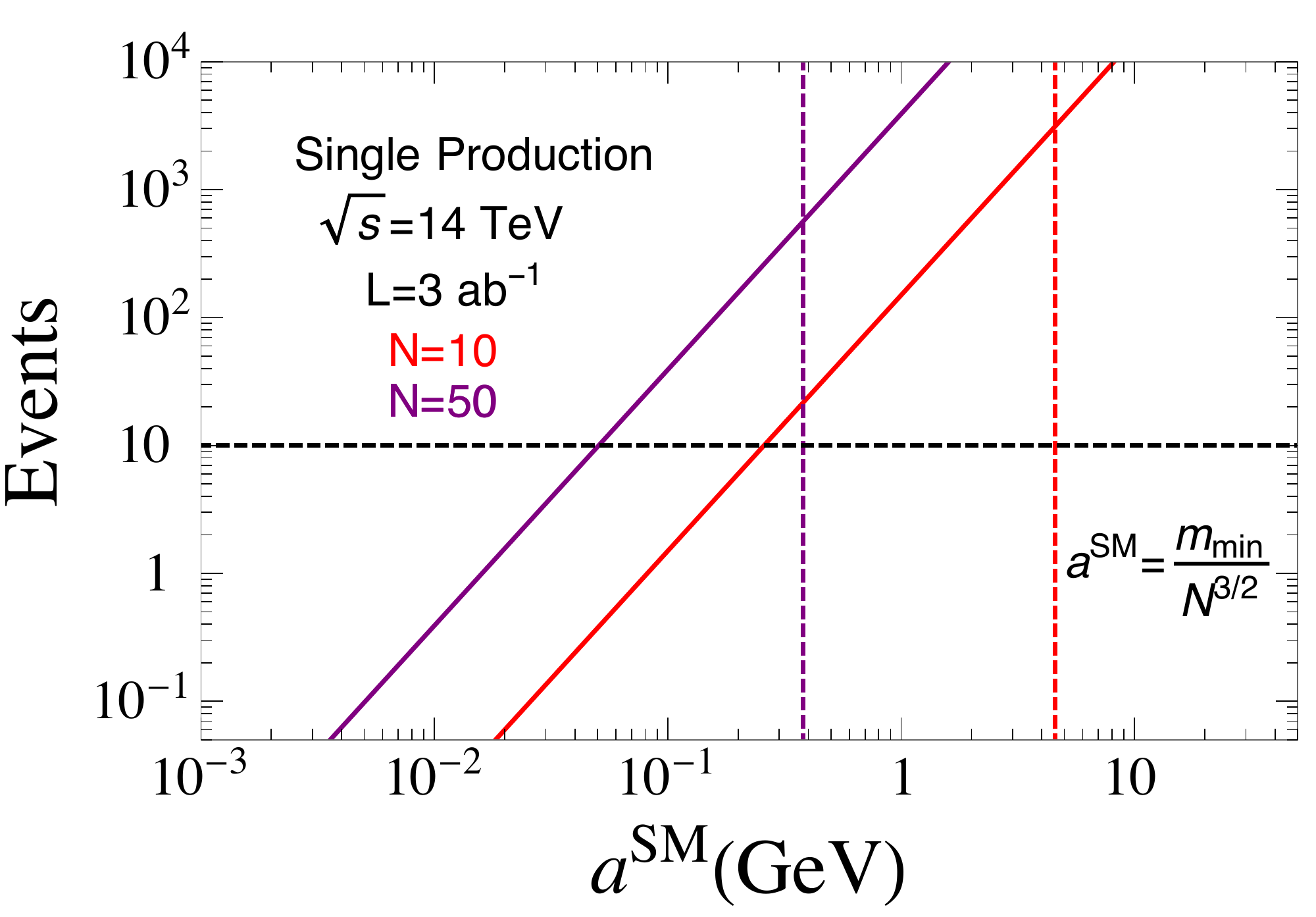}
  \captionsetup{width=0.87\linewidth}
 \caption{Total number of events from pair production (left) and from single production (right) in $3~{\rm ab}^{-1}$ of $\sqrt{s}=14$~TeV LHC data. For pair production the upper end of the lines corresponds to $\lambda^{\rm SM} = 1/N$ and the dotted line indicates where $\lambda^{\rm SM}= 1/N^2$.  For single production the dotted line indicates where $a^{\rm SM} = m_{\rm min}/N^{3/2}$ where $m_{\rm min}$ is the lightest scalar mass.}
  \label{fig:Nevents}
  \end{center}
\end{figure}
%%%%%%%%%%%%%%%%%%%%%%%%%%%%%%%%%%%%%%%%

Before discussing the phenomenology of these three scenarios it is useful to take a look at the scalar production rate at the LHC. We show the total event rate from gluon fusion summed over all pairs of scalars in Fig.~\ref{fig:Nevents} (left) as a fuction of the coupling $\lambda^{\rm SM}$ and summed over all singly produced scalars in Fig.~\ref{fig:Nevents} (right) as a function of the trilinear $a^{\rm SM}$. We use a center-of-mass energy of 14 TeV and an integrated luminosity of $3~{\rm ab}^{-1}$. Cross sections were calculated using Madgraph 5~\cite{Alwall:2011uj} with gluon fusion implemented via the Higgs Effective Theory module~\cite{HiggsEFT}. We have plotted Fig.~\ref{fig:Nevents} for a representative choice of the scalar spectrum drawn at random from a Wigner semicircle distribution in the range of 100 GeV to 600 GeV. The red lines are for $N=10$ and the purple lines are for $N=50$. The figure allows us to conclude that even with our $N$-suppressed couplings we still have a reasonable number of events to work with at the end of the high-luminosity program of the LHC. 
 
%%%%%%%%%%%%%%%%%%%%%%%%%%%%%%%%%%%%%%%%
\begin{figure}[h]
\begin{center}
  \includegraphics[width=0.3\textwidth]{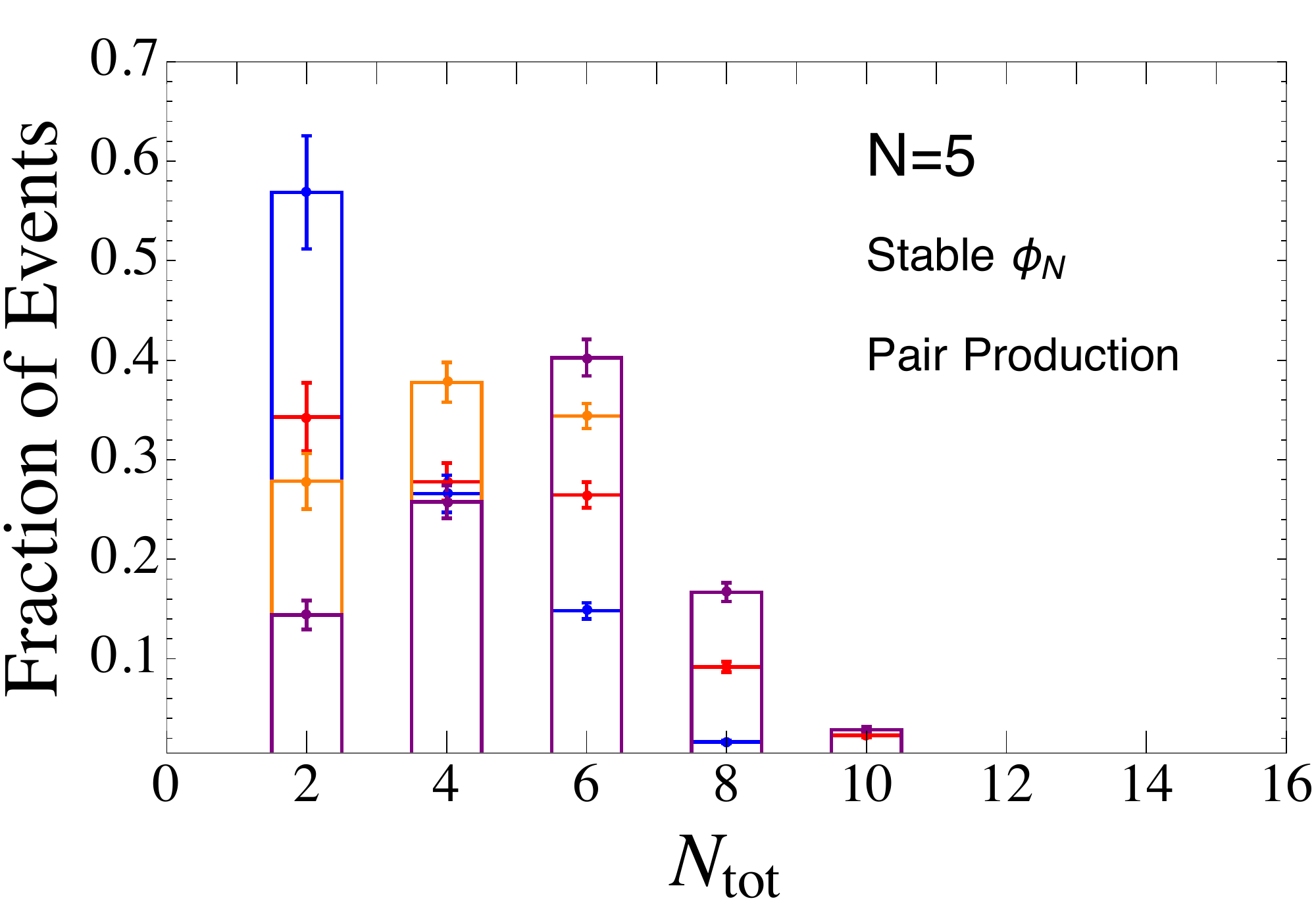}  \quad
  \includegraphics[width=0.3\textwidth]{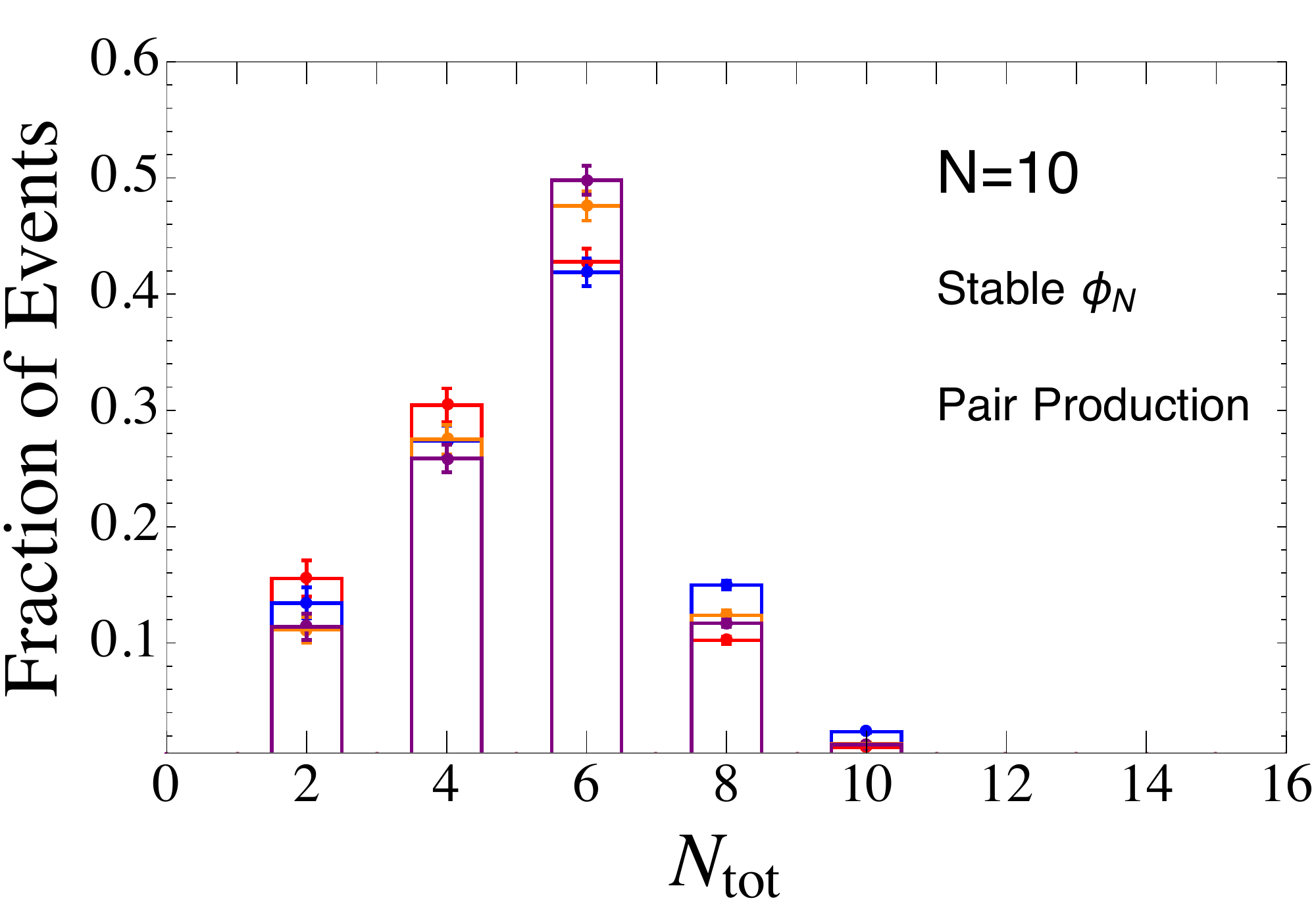} \quad
  \includegraphics[width=0.3\textwidth]{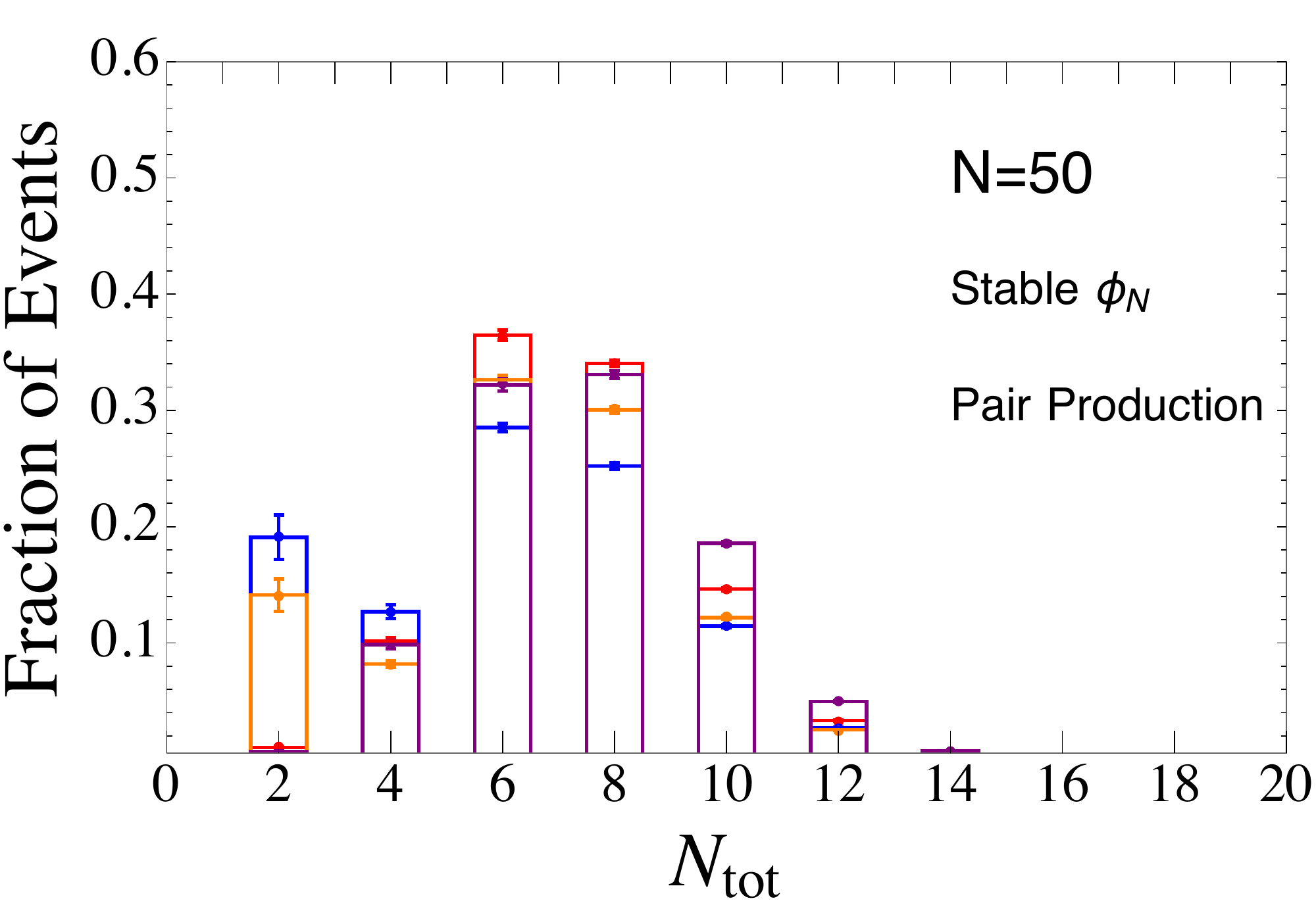}
  \captionsetup{width=0.87\linewidth}
  \caption{Total number of particles in the event in the $Z_2$-symmetric scenario where $a^{\rm SM}=a=0$ in Eqs.~\eqref{eq:scalars} and \eqref{eq:higgscoupling}. Final state particles include two stable scalars. Different colors correspond to different spectra. From left to right: $N$=5, 10, and 50 scalars in the new sector.}
  \label{fig:NtotPair}
\end{center}
\end{figure}
%%%%%%%%%%%%%%%%%%%%%%%%%%%%%%%%%%%%%%%%

%%%%%%%%%%%%%%%%%%%%%%%%%%%%%%%%%%%%%%%%
\begin{figure}[!t]
\begin{center}
\includegraphics[width=0.3\textwidth]{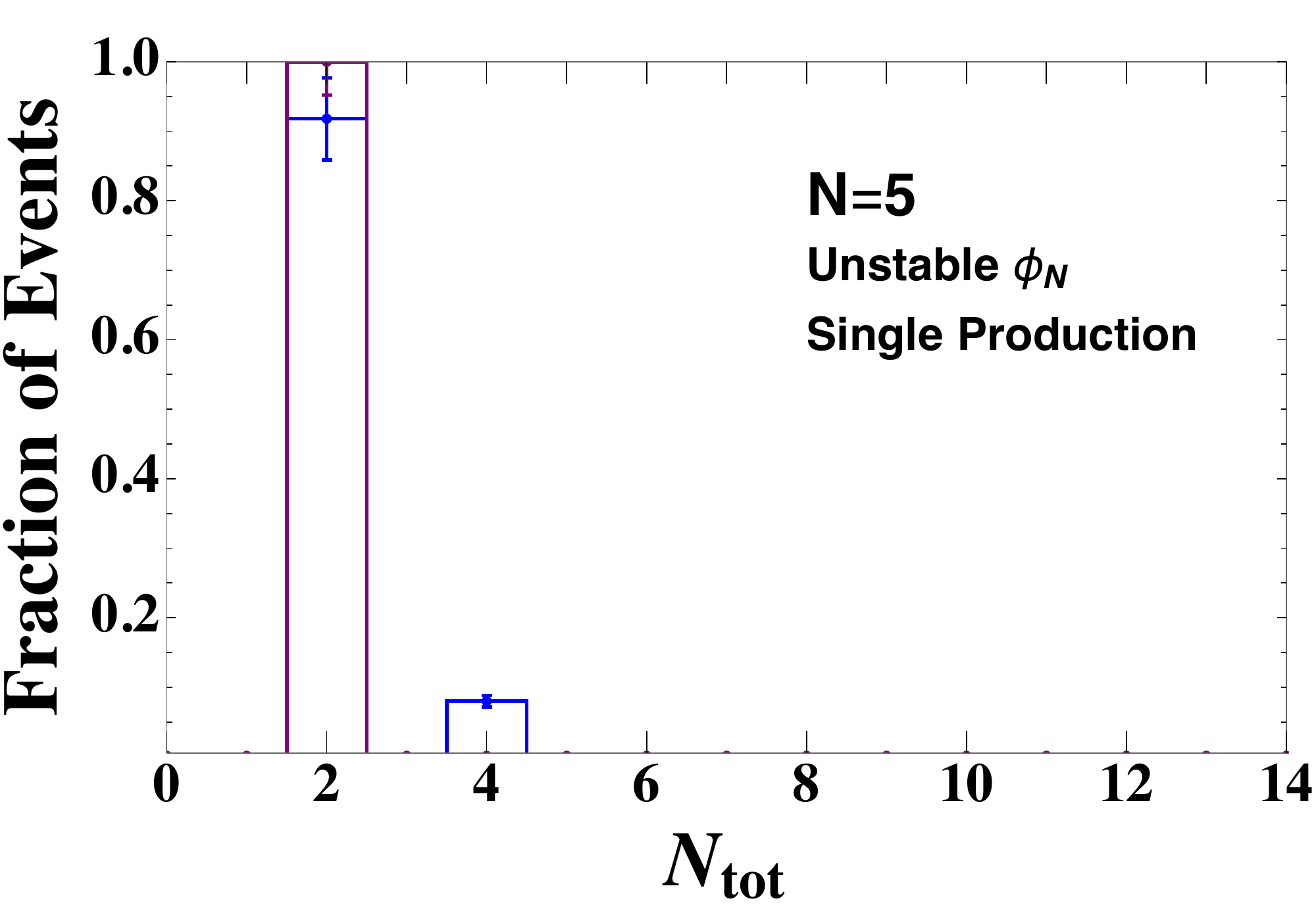}  \quad
\includegraphics[width=0.3\textwidth]{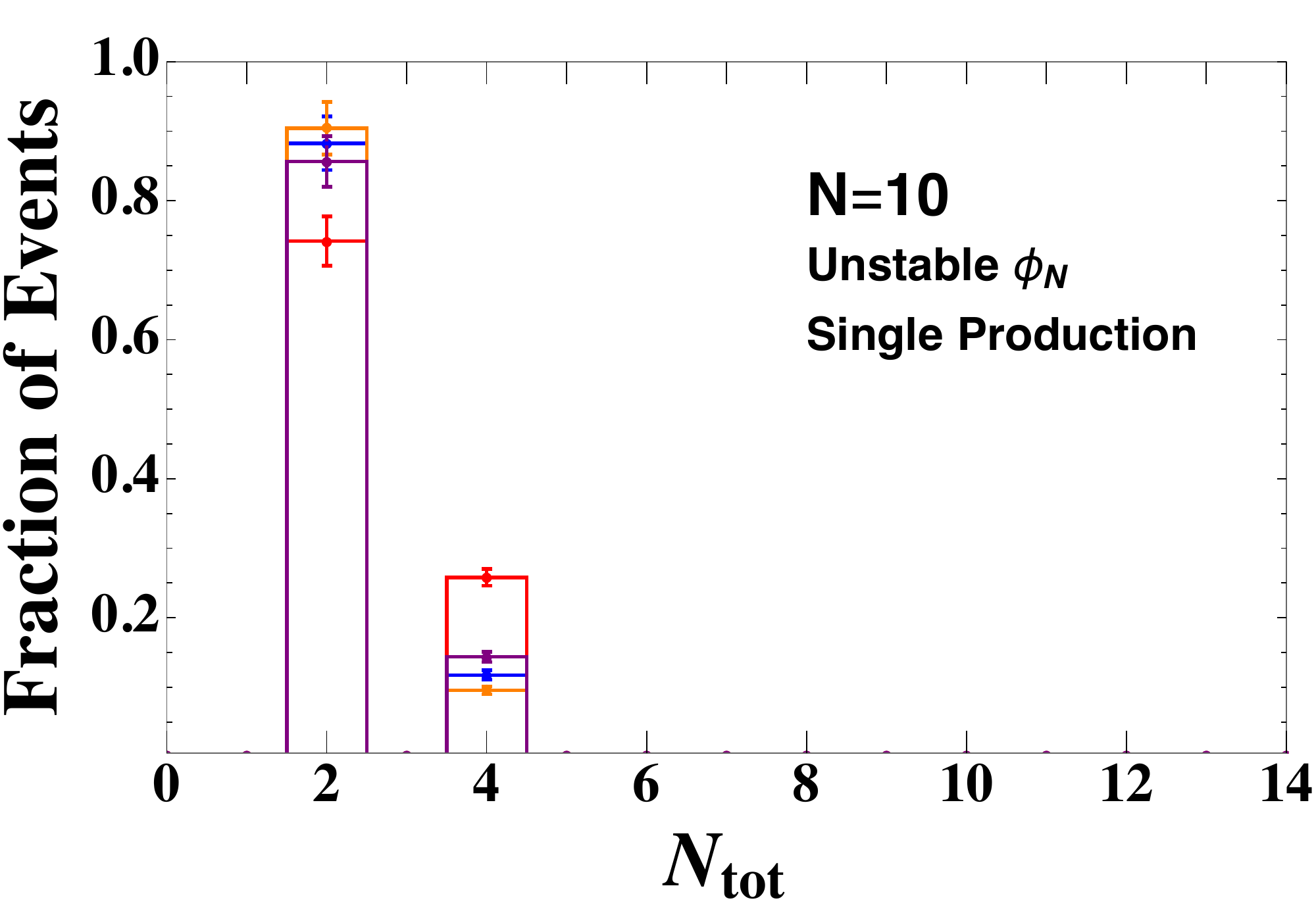} \quad
\includegraphics[width=0.3\textwidth]{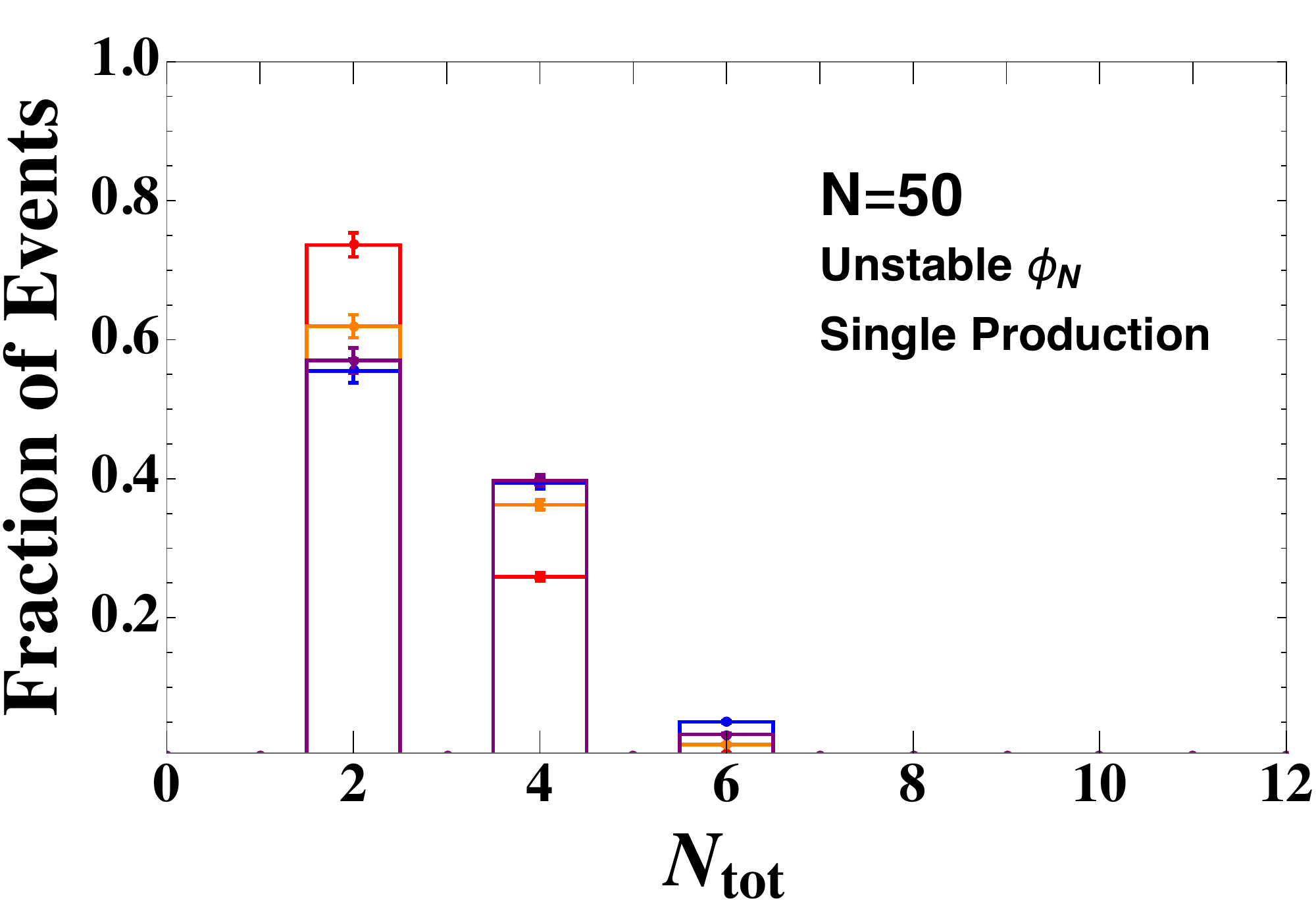}
\captionsetup{width=0.87\linewidth}
 \caption{Total number of particles in the event in the scenario where all couplings in Eqs.~\eqref{eq:scalars} and \eqref{eq:higgscoupling} are present and single production dominates. Different colors correspond to different spectra. From left to right: $N$=5, 10, and 50 scalars in the new sector.}
 \label{fig:NtotSingle}
 \end{center}
\end{figure}
%%%%%%%%%%%%%%%%%%%%%%%%%%%%%%%%%%%%%%%%

%%%%%%%%%%%%%%%%%%%%%%%%%%%%%%%%%%%%%%%%
\begin{figure}[!t]
\begin{center}
\includegraphics[width=0.3\textwidth]{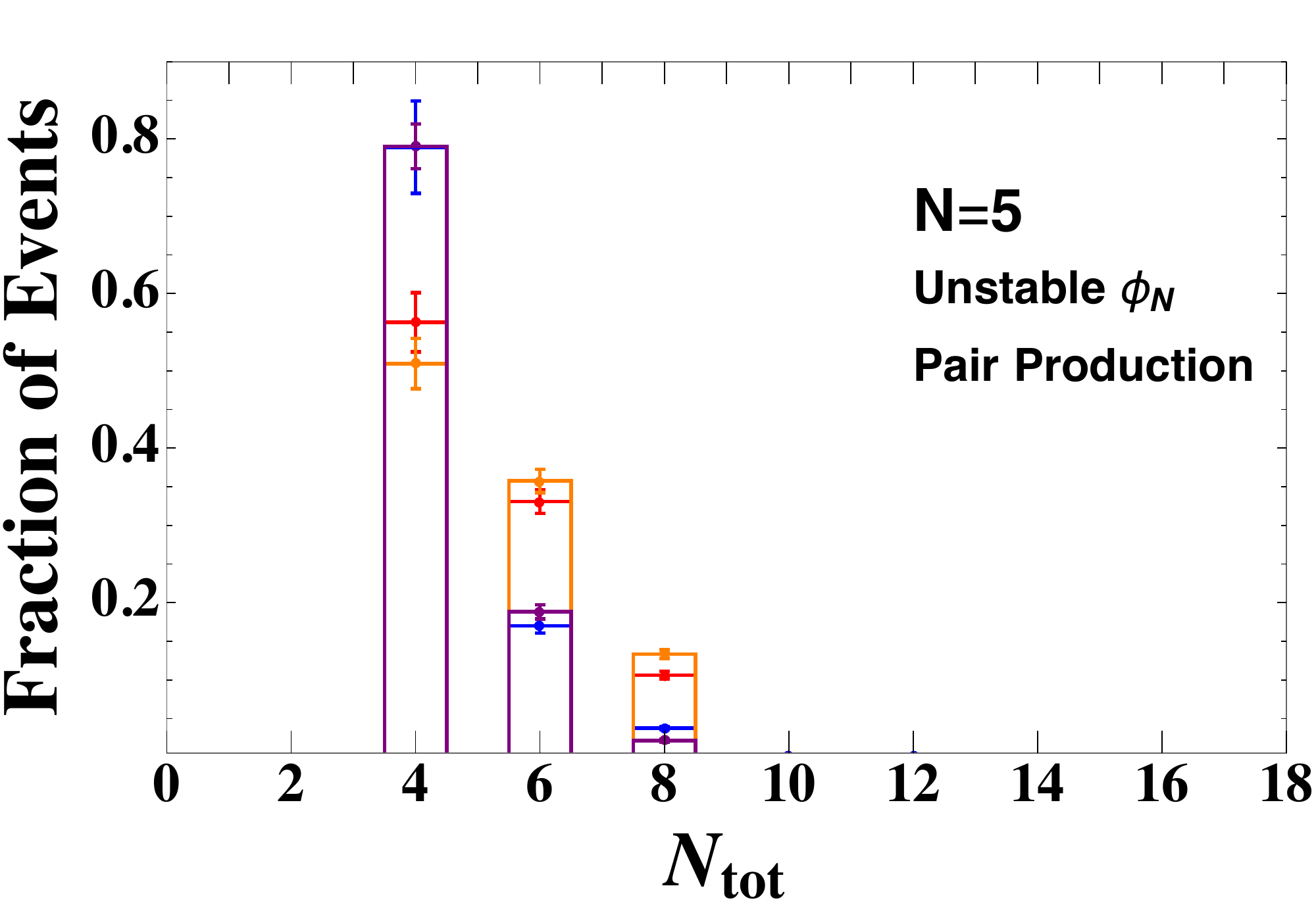} \quad
\includegraphics[width=0.3\textwidth]{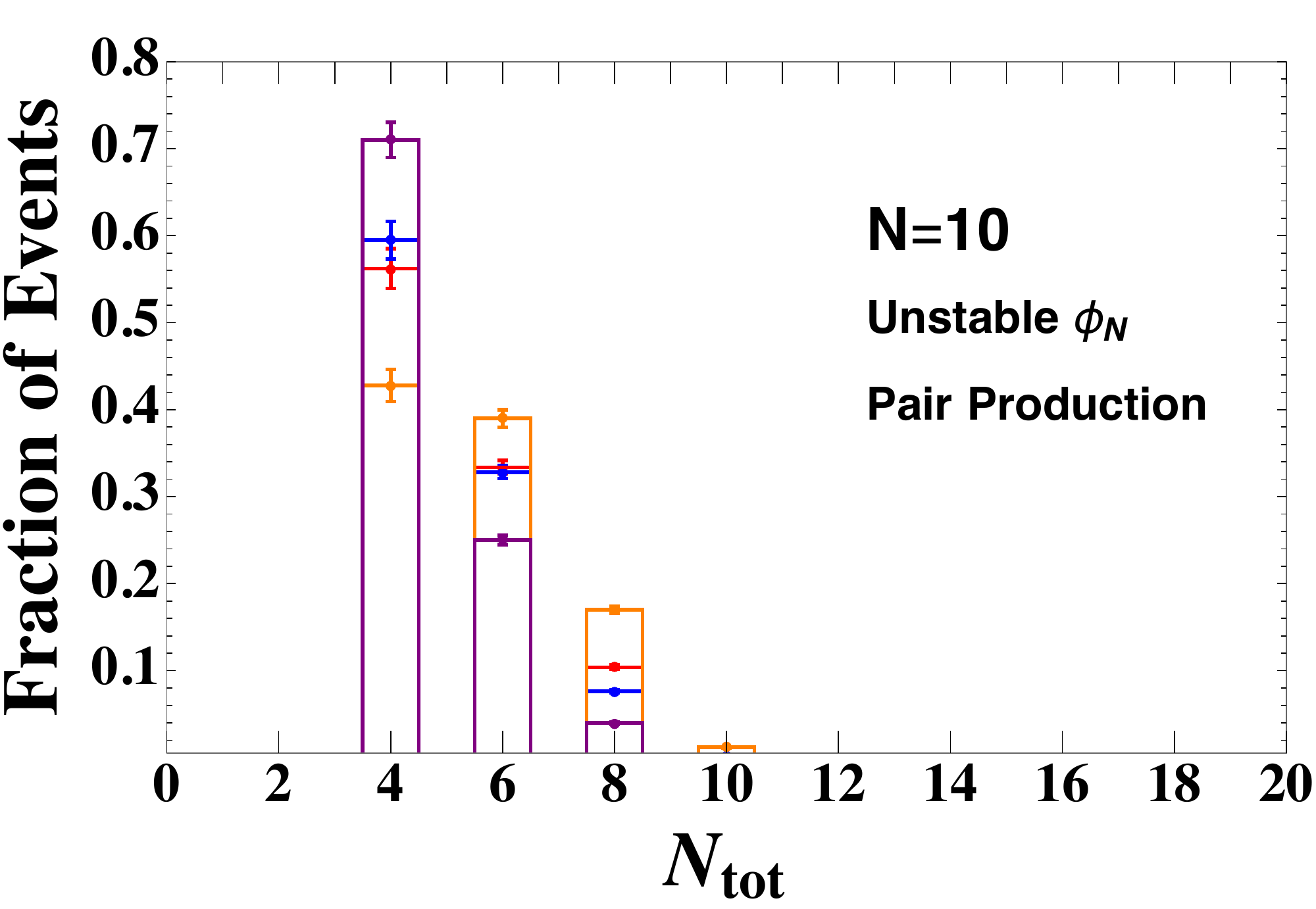} \quad
\includegraphics[width=0.3\textwidth]{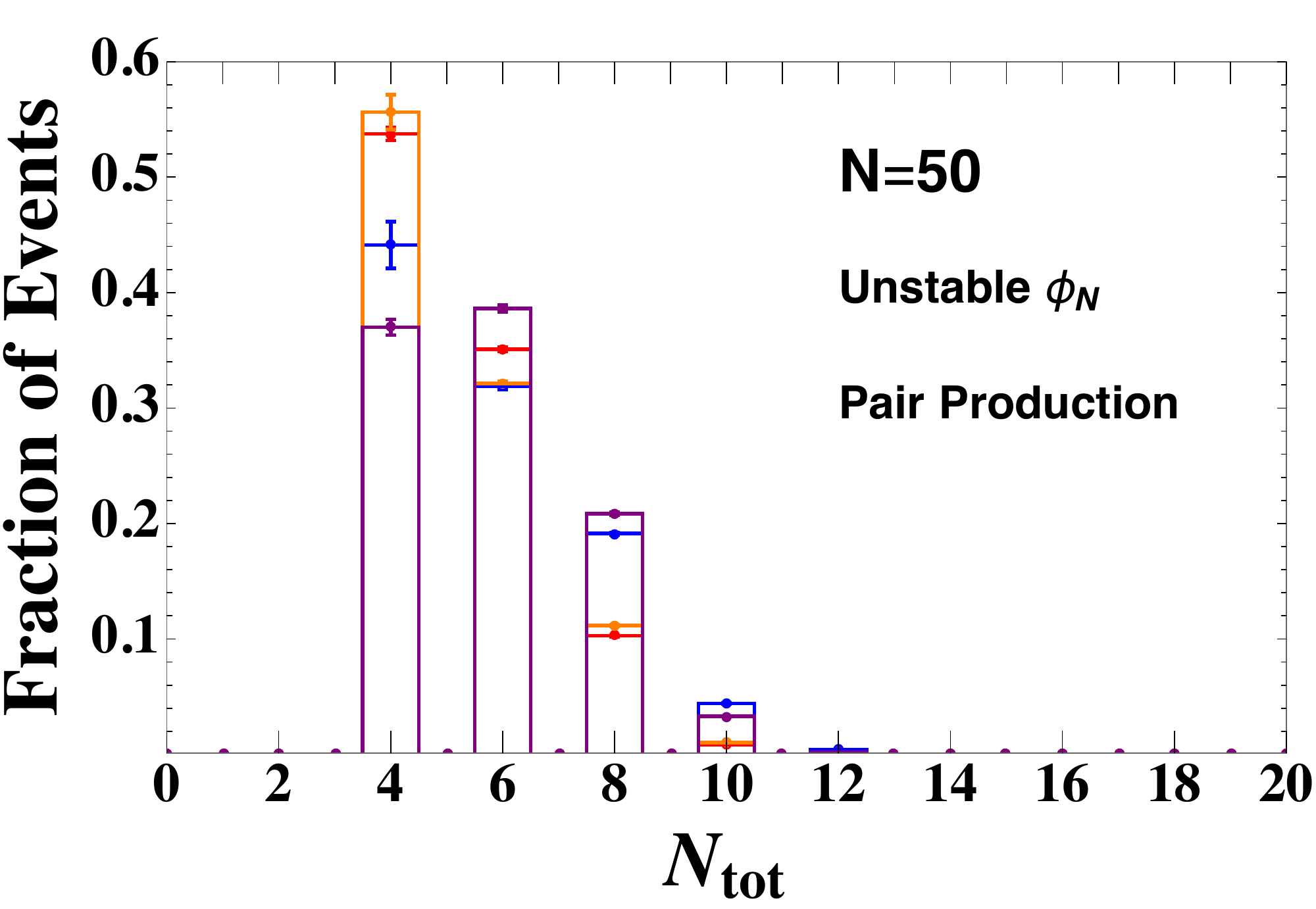}
\captionsetup{width=0.87\linewidth}
 \caption{Total number of particles in the event in the scenario where all couplings in Eqs.~\eqref{eq:scalars} and \eqref{eq:higgscoupling} are present and pair production dominates. Different colors correspond to different spectra. From left to right: $N$=5, 10, and 50 scalars in the new sector.}
 \label{fig:NtotPairT}
 \end{center}
\end{figure}
%%%%%%%%%%%%%%%%%%%%%%%%%%%%%%%%%%%%%%%%

%%%%%%%%%%%%%%%%%%%%%%%%%%%%%%%%%%%%%%%%
\begin{figure}[!t]
  \begin{center}
  \includegraphics[width=0.45\textwidth]{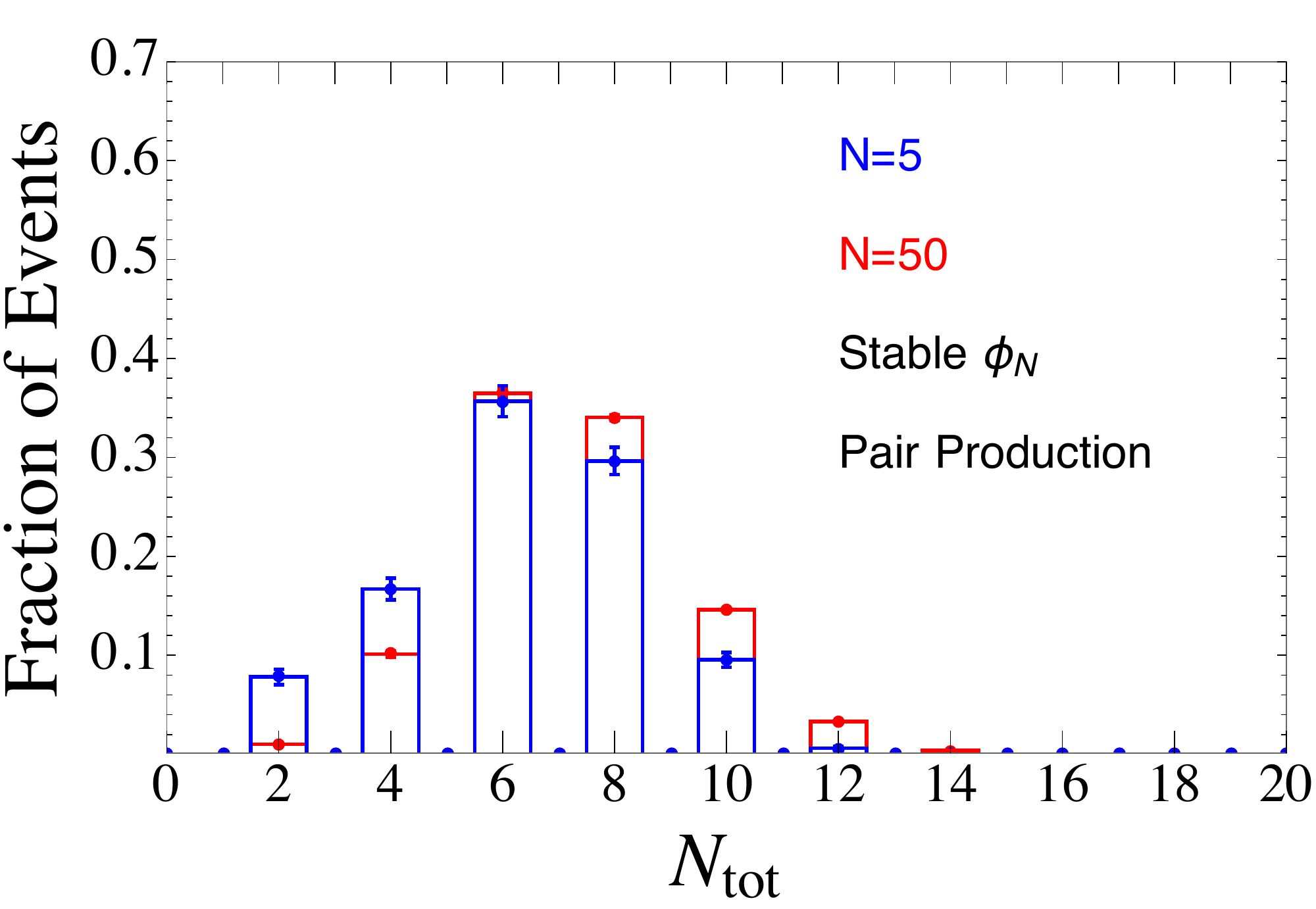}  
  \captionsetup{width=0.87\linewidth}
  \caption{Comparison between 5 states and 50 states with the same average mass splitting. We show the total number of particles in the event for the $Z_2$-symmetric scenario, where $a^{\rm SM}=a=0$ in Eqs.~\eqref{eq:scalars} and \eqref{eq:higgscoupling}. Final state particles include two stable scalars.}
  \label{fig:NtotHD}
  \end{center}
\end{figure}
%%%%%%%%%%%%%%%%%%%%%%%%%%%%%%%%%%%%%%%%

%%%%%%%%%%%%%%%%%%%%%%%%%%%%%%%%%%%%%%%%
\begin{figure}[!t]
\begin{center}
\includegraphics[width=0.3\textwidth]{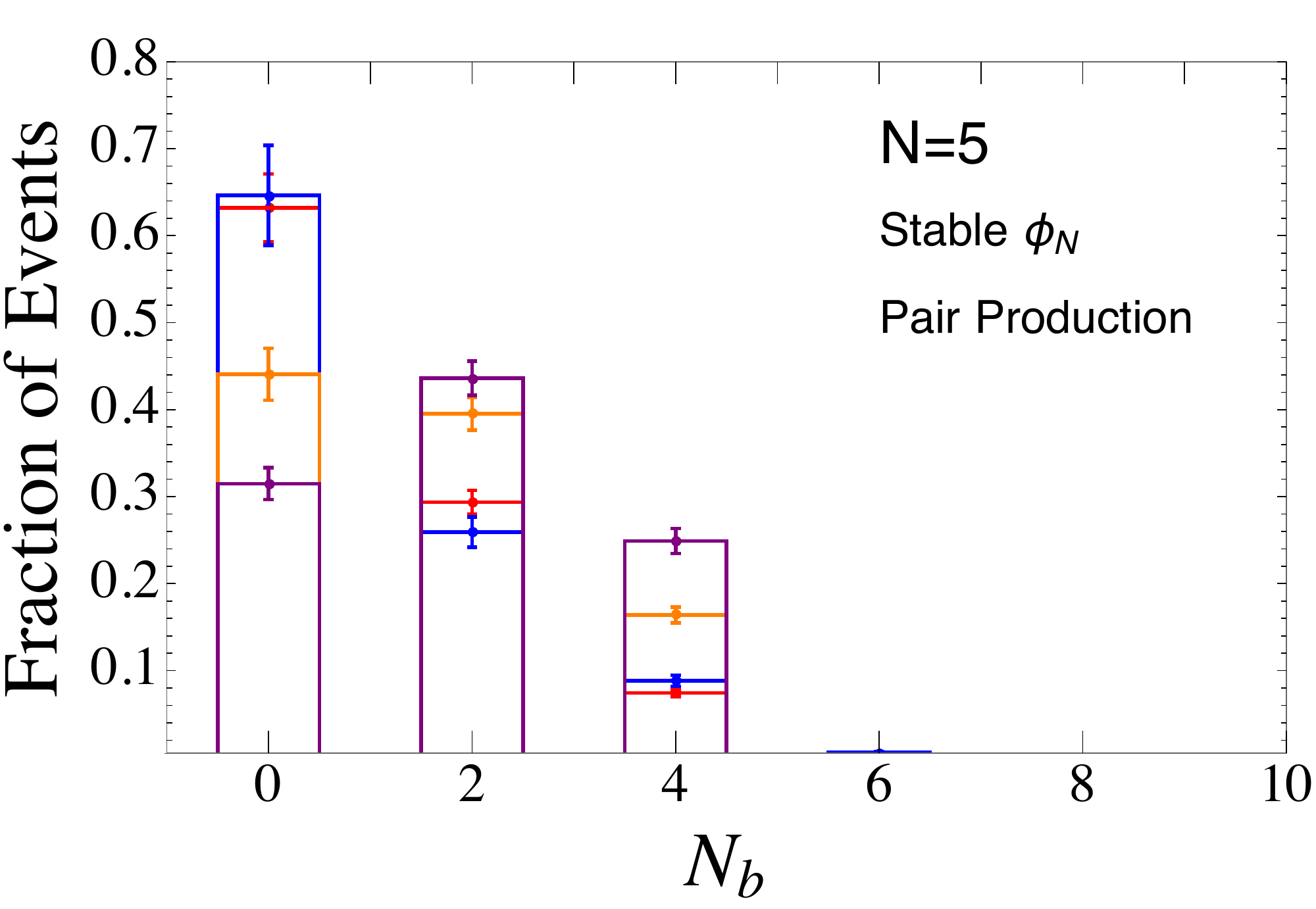}  \quad
\includegraphics[width=0.3\textwidth]{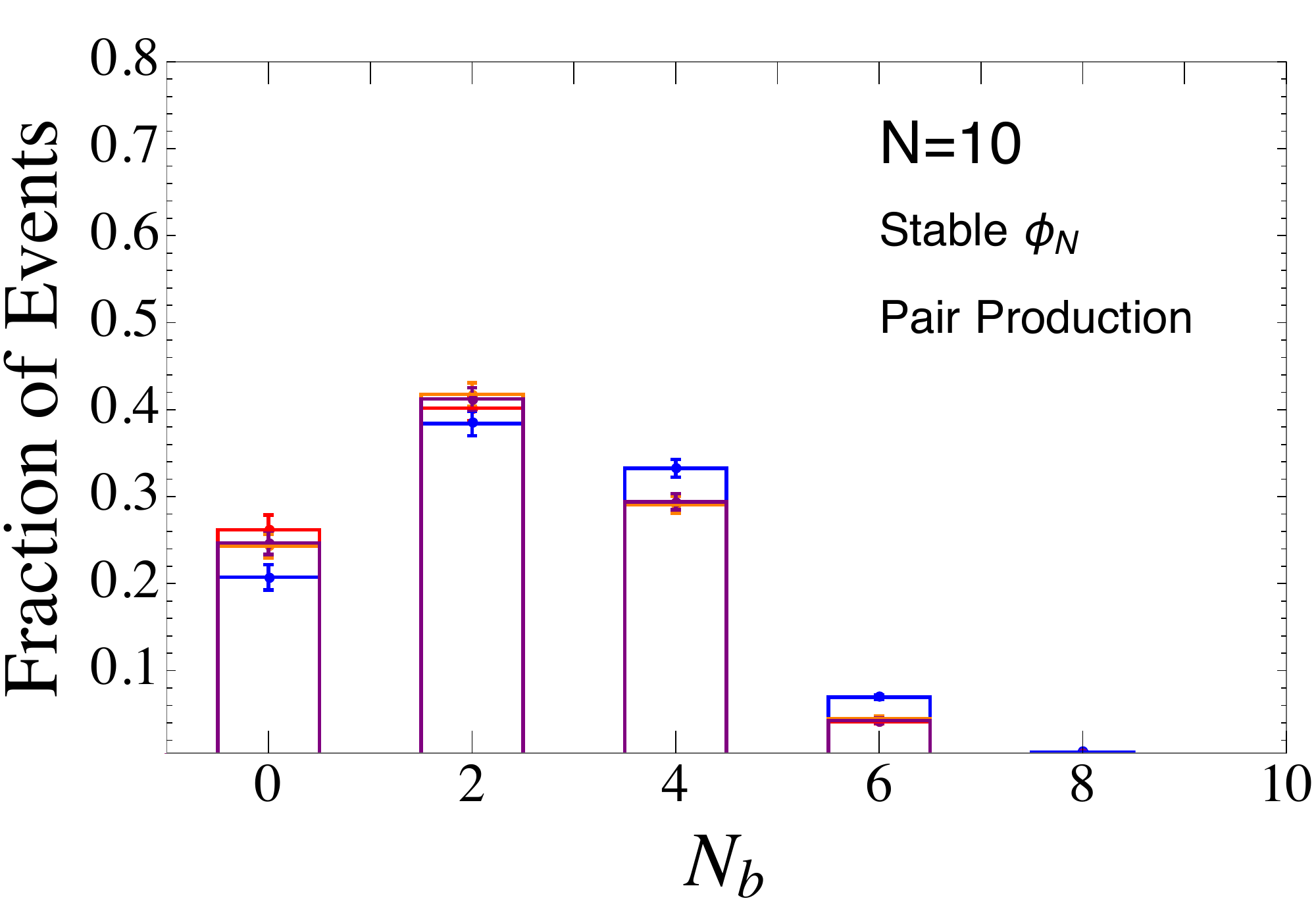} \quad
\includegraphics[width=0.3\textwidth]{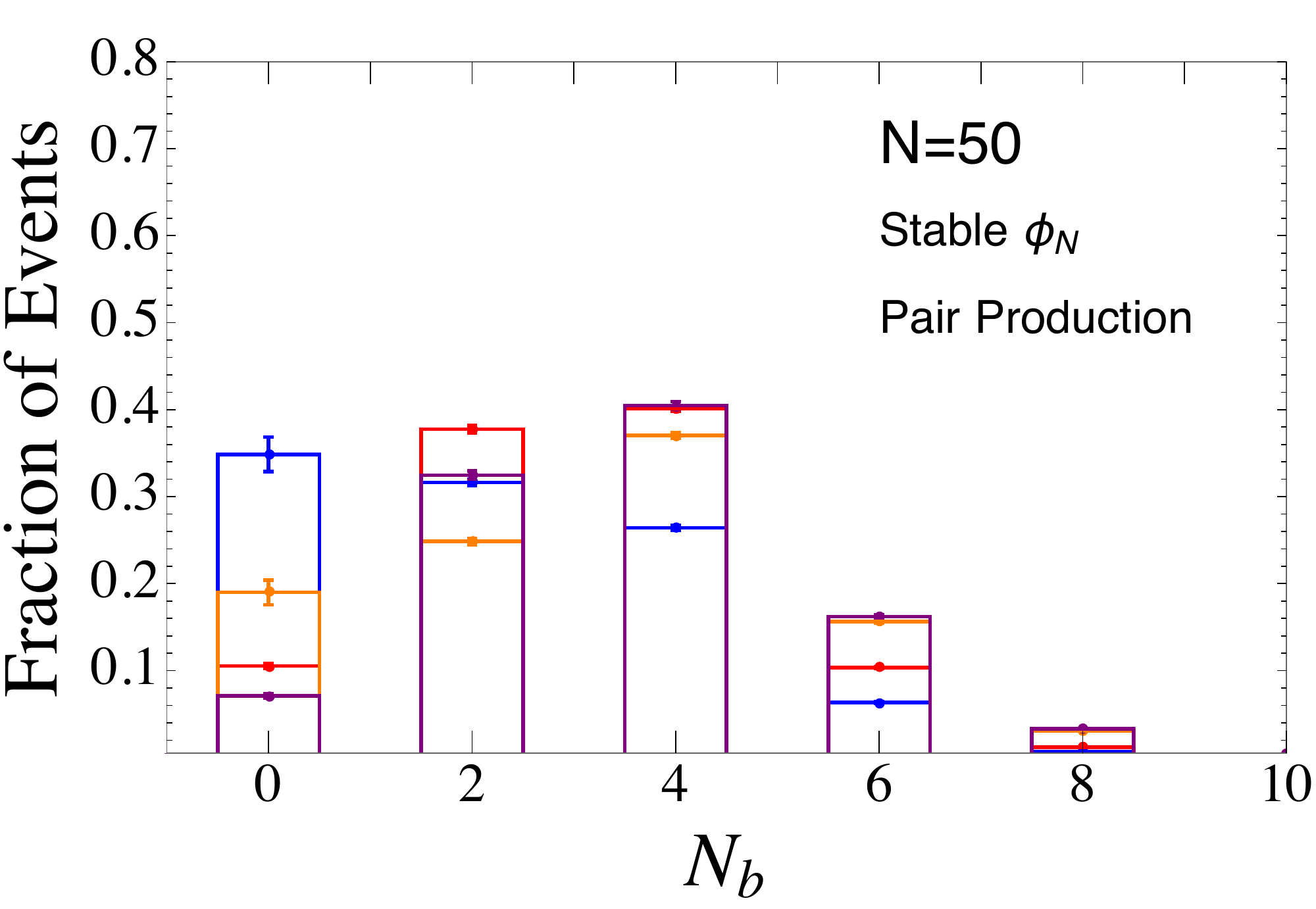}
\captionsetup{width=0.87\linewidth}
 \caption{Total number of $b$-jets in the event for the $Z_2$-symmetric scenario, where $a^{\rm SM}=a=0$ in Eqs.~\eqref{eq:scalars} and \eqref{eq:higgscoupling}. Different colors correspond to different spectra. From left to right: $N$=5, 10, and 50 scalars in the new sector.}
 \label{fig:NBPair}
 \end{center}
\end{figure}
%%%%%%%%%%%%%%%%%%%%%%%%%%%%%%%%%%%%%%%%

%%%%%%%%%%%%%%%%%%%%%%%%%%%%%%%%%%%%%%%%
\begin{figure}[!t]
\begin{center}
\includegraphics[width=0.3\textwidth]{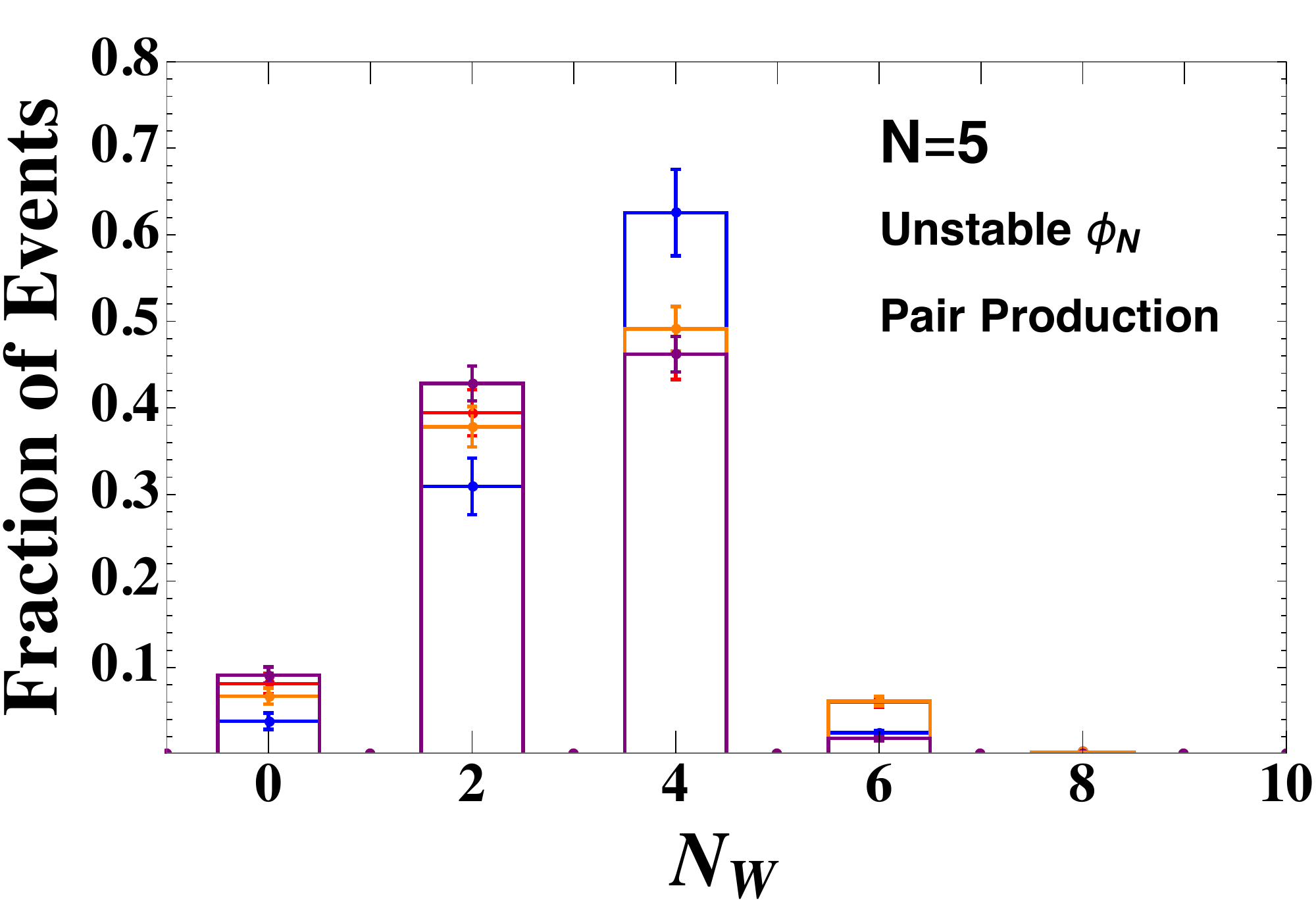}  \quad
\includegraphics[width=0.3\textwidth]{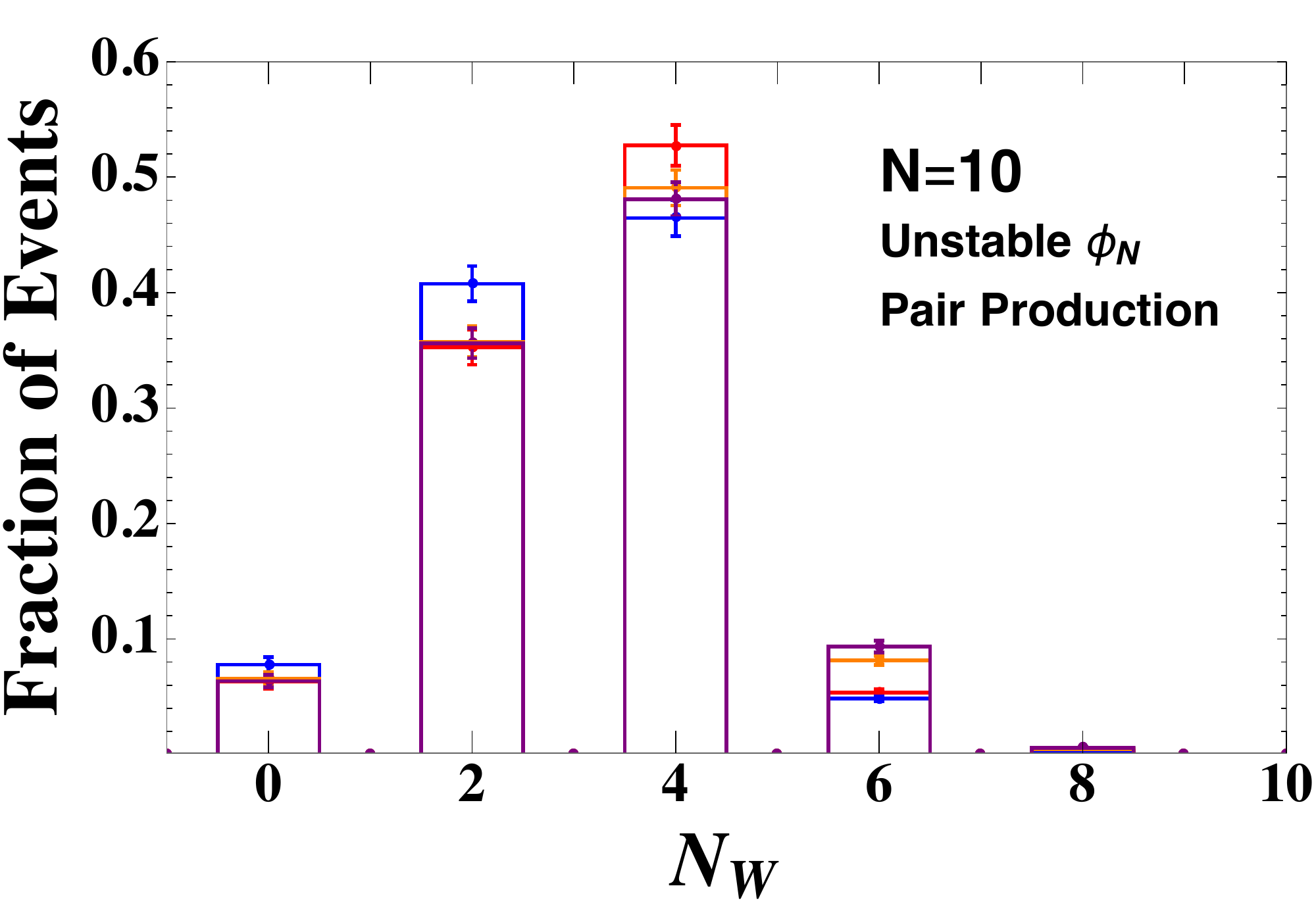} \quad
\includegraphics[width=0.3\textwidth]{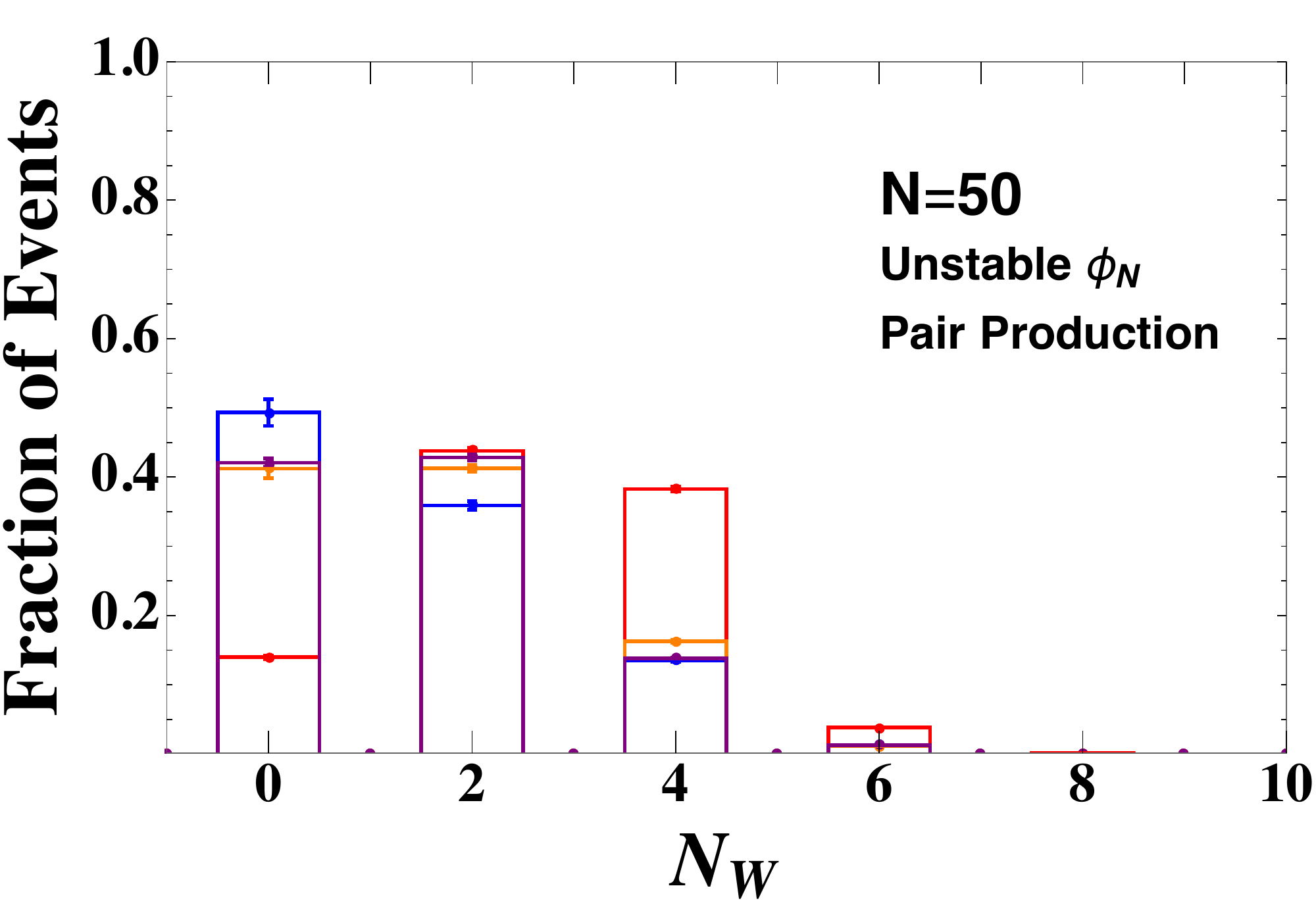}
\captionsetup{width=0.87\linewidth}
 \caption{Total number of $W$ bosons in the event in the scenario where all couplings in Eqs.~\eqref{eq:scalars} and \eqref{eq:higgscoupling} are present and pair production dominates. Different colors correspond to different spectra. From left to right: $N$=5, 10, and 50 scalars in the new sector.}
 \label{fig:NWPairT}
 \end{center}
\end{figure}
%%%%%%%%%%%%%%%%%%%%%%%%%%%%%%%%%%%%%%%%

To analyze the phenomenological features of the model we start by showing the total number of particles $N_{\rm tot}$ per event. In Fig.~\ref{fig:NtotPair} we show $N_{\rm tot}$ for $N=5$, $N=10$, and $N=50$ scalars in the $Z_2$-symmetric case.  In Fig.~\ref{fig:NtotSingle} we show the same for the second case, where single production dominates, while in Fig.~\ref{fig:NtotPairT} we show $N_{\rm tot}$ for the last scenario that is nearly $Z_2$-symmetric, but has small trilinears that allow the lightest scalar to decay. The different colors correspond to a different randomly-generated mass spectrum, so the four histograms in each figure represent the variation that we expect from randomness in the mass matrix.

Especially at small $N$, the variation between different spectra is primarily a function of the gap between the lightest state $\phi_N$ and the states closest in mass to it.  When there is a large gap between $\phi_N$ and the next state, events with $N_{\rm tot}=2$ are favored because $\phi_N \phi_N$ production dominates the overall rate.  Spectra for which the $N_{\rm tot}=2$ bin is small have states that are near in mass to $\phi_N$ so that their production rate is comparable to that of $\phi_N$.  Furthermore, smaller mass gaps near the bottom of the spectrum mean that heavy scalars have a very similar probability to decay to any of the light scalars favoring longer decay chains.  If $\phi_N$ is much lighter, phase space will instead favor a direct decay to the bottom of the spectrum. 

This discussion leads us to identify the main qualitative feature common to all three regions of parameter space: increasing the density of states in a fixed mass interval, long decay chains become more common, giving rise to higher multiplicity final states. This emerges clearly from Fig.~\ref{fig:NtotHD} where we compare one of the spectra of Fig.~\ref{fig:NtotPair} for $N=50$ with $N=5$ scalars in the same scenario, but distributed over a much smaller mass range: from 200 to 250 GeV. The two spectra have the same average mass splitting between neighboring states and very similar final state multiplicities.

\begin{figure}[!t]
  \begin{center}
  \includegraphics[width=0.45\textwidth]{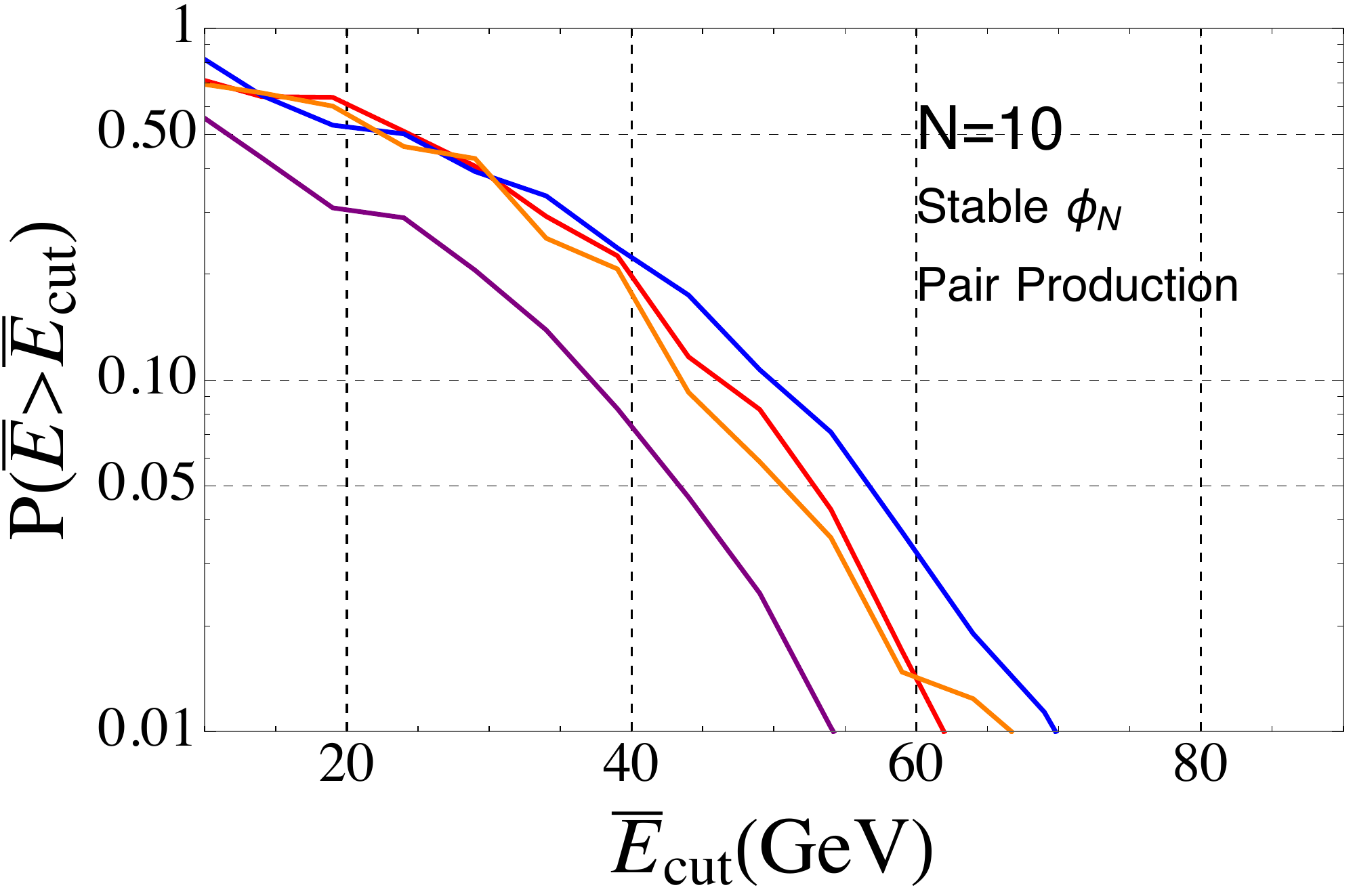} \quad
  \includegraphics[width=0.45\textwidth]{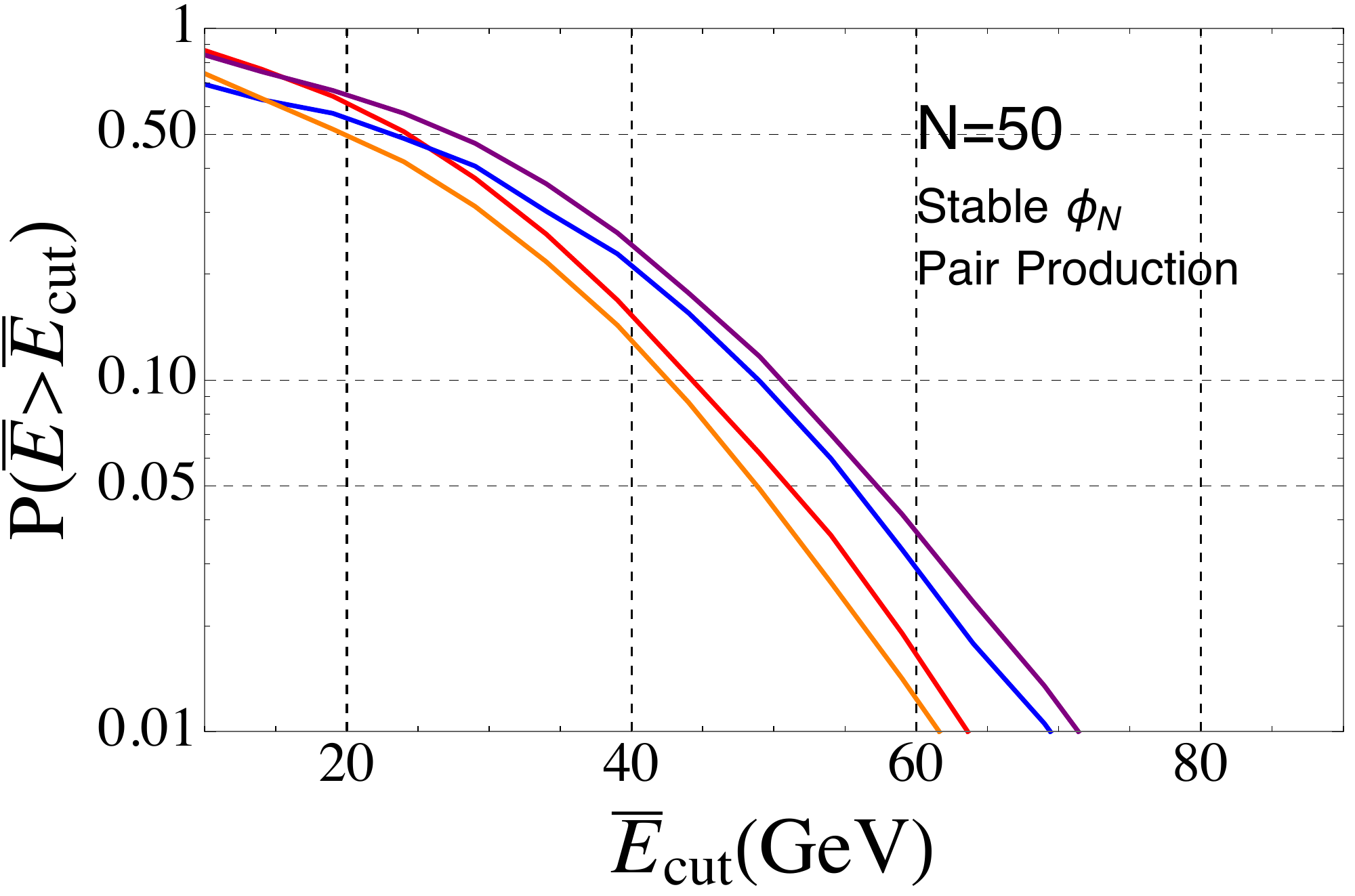}
  \captionsetup{width=0.87\linewidth}
  \caption{Fraction of events with average energy per particle above $\overline E_{\rm cut}$. The plots were made in the $Z_2$-symmetric scenario where $a^{\rm SM}=a=0$ in Eqs.~\eqref{eq:scalars} and \eqref{eq:higgscoupling}. Left: 10 scalars, right: 50 scalars.}
  \label{fig:AvEPair}
  \end{center}
\end{figure}

\begin{figure}[!t]
  \begin{center}
  \includegraphics[width=0.45\textwidth]{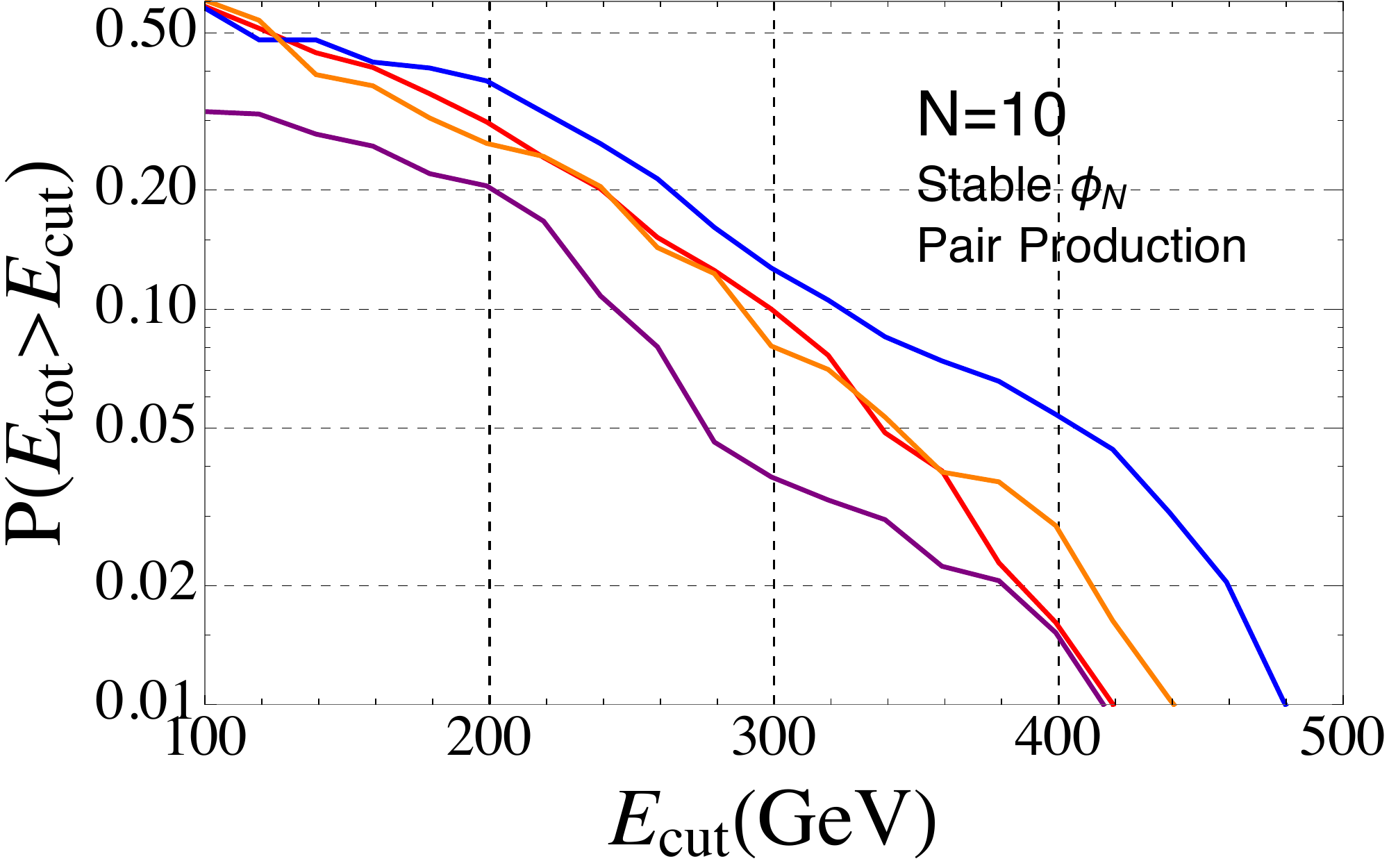} \quad
  \includegraphics[width=0.45\textwidth]{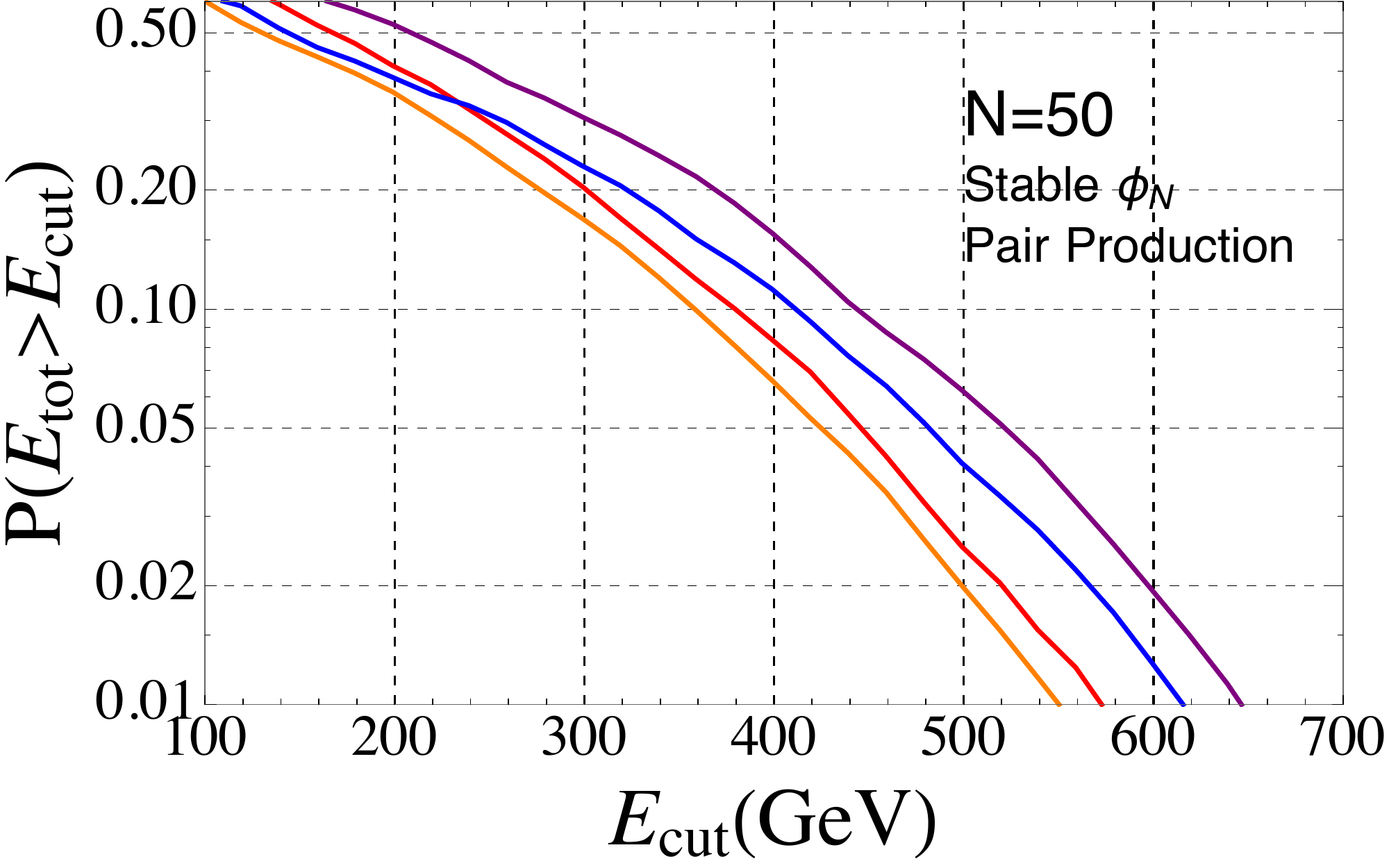}
  \captionsetup{width=0.87\linewidth}
  \caption{Fraction of events with total energy (including missing energy) above $E_{\rm cut}$. The plots were made in the $Z_2$-symmetric scenario where $a^{\rm SM}=a=0$ in Eqs.~\eqref{eq:scalars} and \eqref{eq:higgscoupling}. Left: 10 scalars, right: 50 scalars.}
  \label{fig:TotEPair}
  \end{center}
\end{figure}

From Figs.~\ref{fig:NtotPair},~\ref{fig:NtotSingle}, and~\ref{fig:NtotPairT} another general aspect of the phenomenology of the model emerges clearly: when single production dominates traditional resonant searches for pairs of SM objects are still the main avenue for discovery. However when pair production dominates the main signature consists in four or more particles in the final state. In the $Z_2$-symmetric case these particles are mostly soft $b$-quark jets. This is shown in Fig.~\ref{fig:NBPair} and Fig.~\ref{fig:AvEPair}. The first figure counts the number of $b$-quarks in the event, while the second one shows the fraction of events with average energy per particle above a certain threshold for $N=10$ and $N=50$ scalars. From Fig.~\ref{fig:AvEPair} it is clear that our final states are extremely challenging for traditional low-multiplicity triggers and even the total energy in the event, shown in Fig.~\ref{fig:TotEPair} is not a good handle. A quantitative discussion of trigger thresholds goes beyond the scope of this work, but it is clear from our results that low-threshold, high-multiplicity triggers are well motivated in this scenario. Note also that in this scenario the lightest scalar $\phi_N$ is stable and we have always a small amount of missing energy in the event. It is too small for triggering purposes, but it can be used as a handle to identify this model.

%%%%%%%%%%%%%%%%%%%%%%%%%%%%%%%%%%%%%%%%
%%%%%%%%%%%%%%%%%%%%%%%%%%%%%%%%%%%%%%%%

%%%%%%%%%%%%%%%%%%%%%%%%%%%%%%%%%%%%%%%%
%%%%%%%%%%%%%%%%%%%%%%%%%%%%%%%%%%%%%%%%

%%%%%%%%%%%%%%%%%%%%%%%%%%%%%%%%%%%%%%%%
\begin{figure}[!t]
  \begin{center}
  \includegraphics[width=0.45\textwidth]{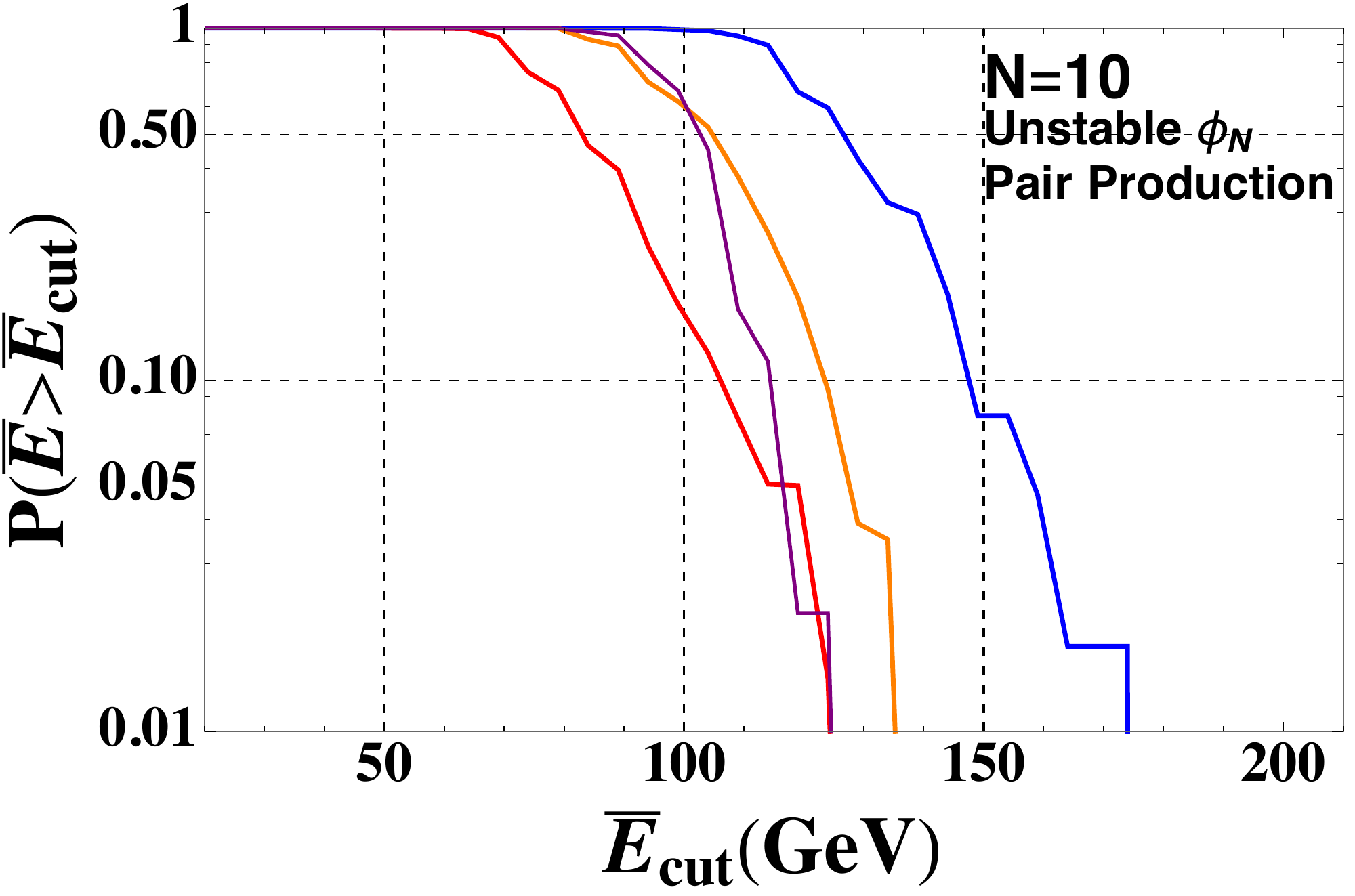} \quad
  \includegraphics[width=0.45\textwidth]{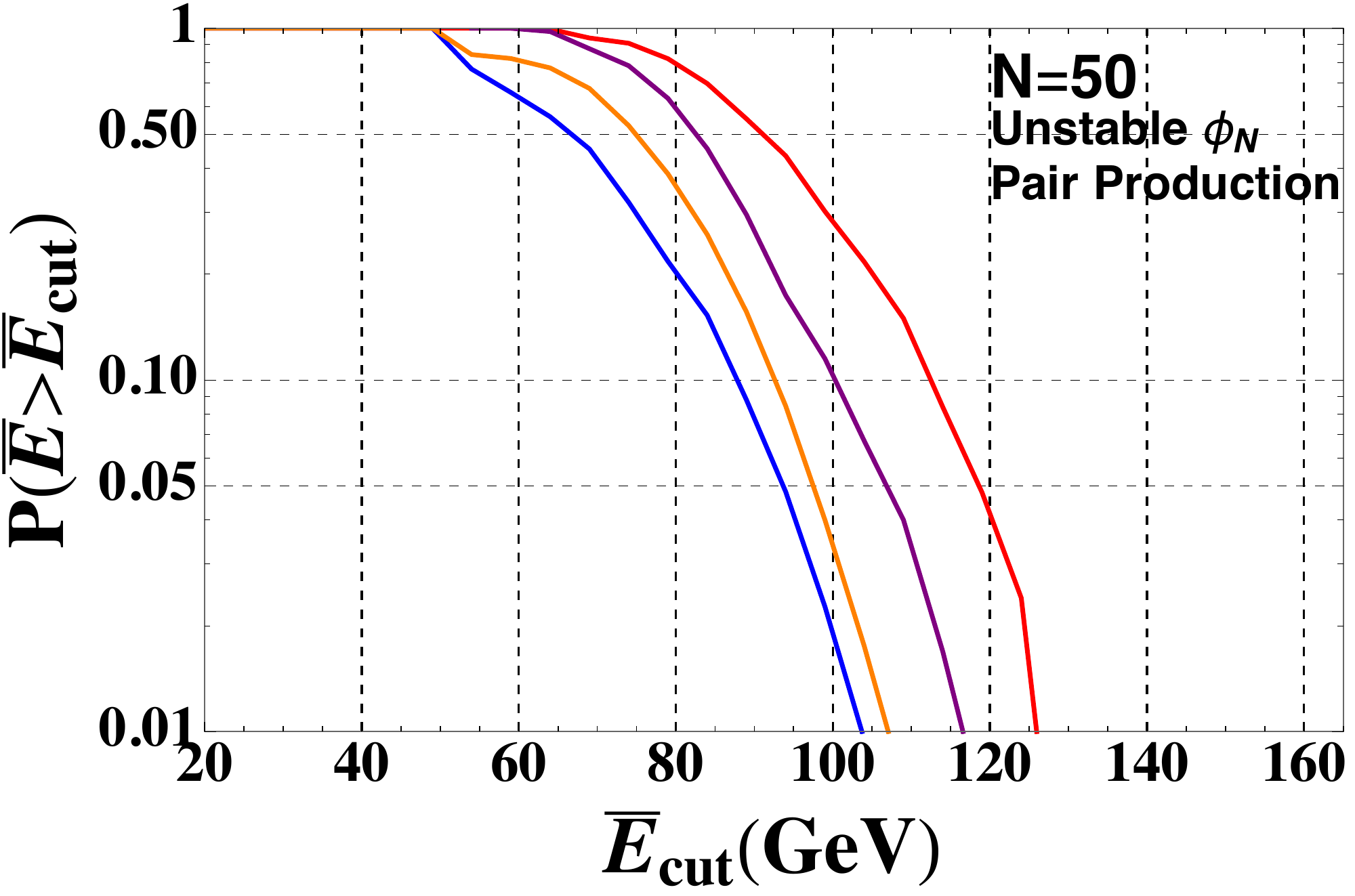}
  \captionsetup{width=0.87\linewidth}
  \caption{Fraction of events with average energy per particle above $\overline E_{\rm cut}$. The plots were made in the scenario where all couplings in Eqs.~\eqref{eq:scalars} and \eqref{eq:higgscoupling} are present and pair production dominates. Left: 10 scalars, right: 50 scalars.}
  \label{fig:AvEPairT}
  \end{center}
\end{figure}
%%%%%%%%%%%%%%%%%%%%%%%%%%%%%%%%%%%%%%%%

%%%%%%%%%%%%%%%%%%%%%%%%%%%%%%%%%%%%%%%%
\begin{figure}[!t]
  \begin{center}
  \includegraphics[width=0.45\textwidth]{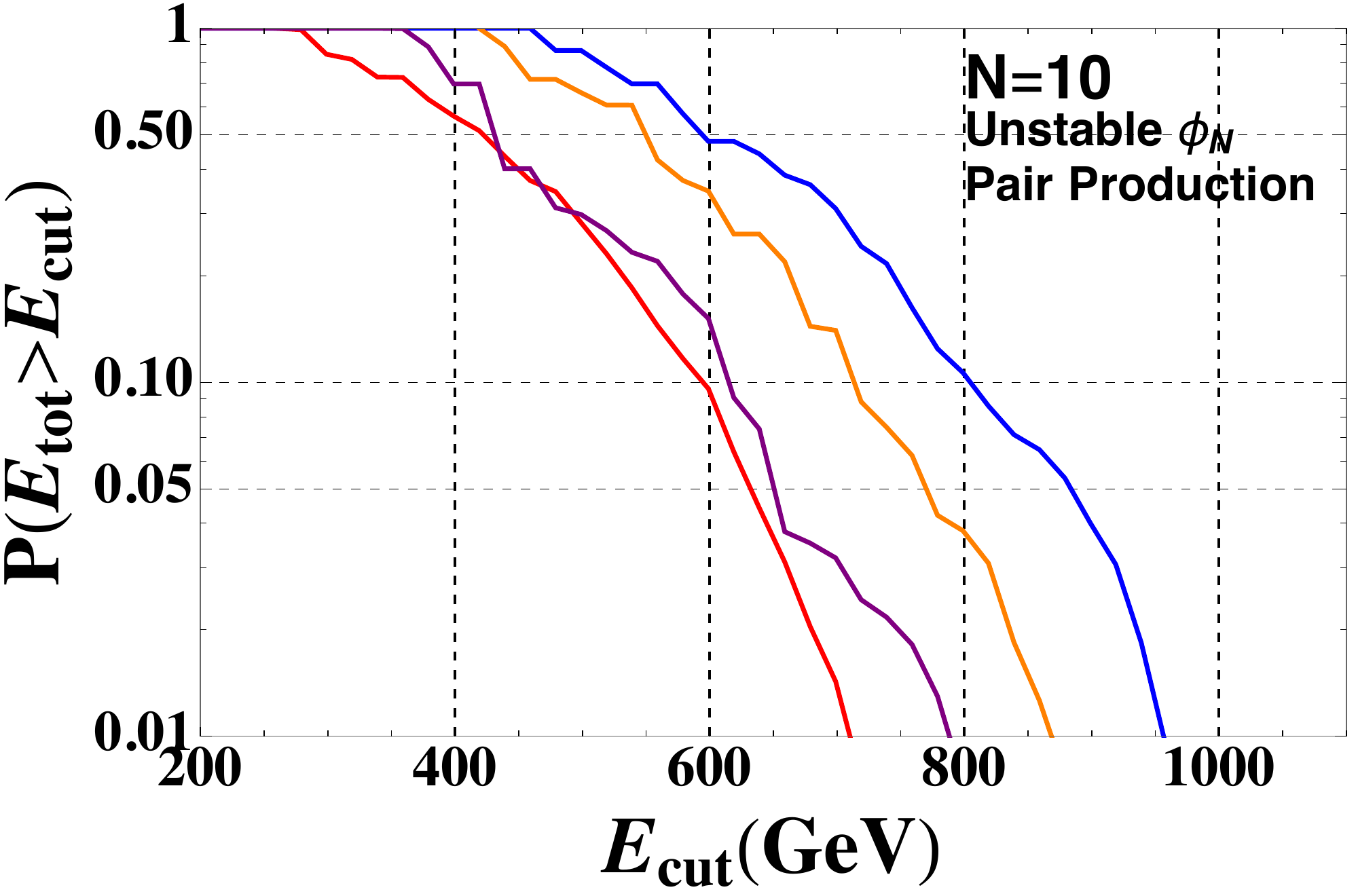} \quad
  \includegraphics[width=0.45\textwidth]{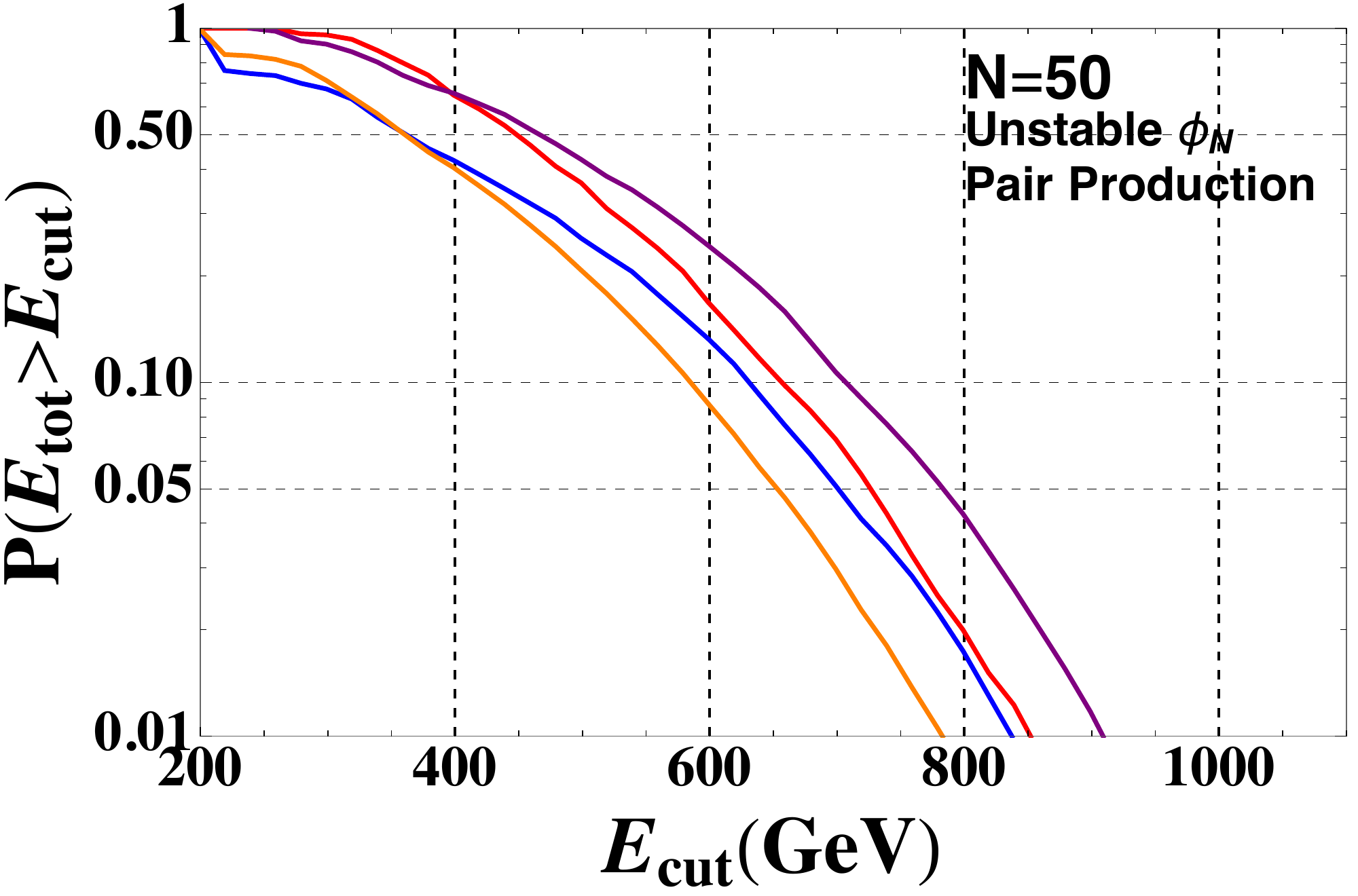}
  \captionsetup{width=0.87\linewidth}
  \caption{Fraction of events with total energy (including missing energy) above $E_{\rm cut}$. The plots were made in the scenario where all couplings in Eqs.~\eqref{eq:scalars} and \eqref{eq:higgscoupling} are present and pair production dominates.  Left: 10 scalars, right: 50 scalars.}
  \label{fig:TotEPairT}
  \end{center}
\end{figure}
%%%%%%%%%%%%%%%%%%%%%%%%%%%%%%%%%%%%%%%%

Let us now turn to the last scenario, where pair production dominates but $\phi_N$ can decay to a pair of SM particles.  Interestingly for $N\lesssim 10$ most of the events contain at least four $W$s, as shown in Fig.~\ref{fig:NWPairT}. At larger $N$ $b$-quarks are still the dominant species of SM particles.  In this scenario $\phi_N$ decays via its mixing with the Higgs. So the large $W$ multiplicity is due to the mass of the lightest state in the spectrum, that for small $N$ is usually larger than the Higgs mass (due to the shape of the Wigner semicircle distribution). The $WW$ decay width turns on rapidly for $m_W \lesssim m_\phi \lesssim 2 m_W$. On the contrary at larger $N$ when we start to populate the entire Wigner distribution, we always find a scalar with mass comparable or smaller than $m_h$ for which decays to SM quarks dominate. From the point of view of triggering the total energy in the event still does not offer a very good handle as shown in Fig.~\ref{fig:TotEPairT}. However at small $N$ the average energy per particle (shown in Fig.~\ref{fig:AvEPairT}) is more than enough for leptonic and some multijet triggers. 

To summarize, the phenomenology of the model is very rich, ranging from scenarios where traditional resonant searches capture the bulk of the events to cases where long decay chains with multiple $b$s or $W$s are the most common signatures. In general the total energy and missing energy in the event cannot be used for triggering and high-multiplicity triggers are motivated. In this section we took the mass range of the scalars between 100 and 600 GeV. It would be interesting to explore different mass ranges. Going to larger masses would boost the total energy in the event potentially changing our qualitative conclusions on triggering. However this can be done only at the price of considerably reducing the production rate and would not make these new sectors less elusive. Going to lower masses would make the new scalars even harder to trigger on or move them to kinematical regions better explored outside of the domain of hadron colliders. However following the agnostic approach outlined in the introduction it would be worth to consider also much lighter sectors and a completely different set of experiments. We leave this exploration to future work.

%**************** Section ***********************************
\section{Fermionic Hidden Sectors}
\label{sec:fermions}
%*************************************************************

In this paper we have chosen to focus on models with scalars and explore thoroughly their mass matrices. However the same ideas could be realized in models containing a large number of fermions or a mixture of particles with different spins. In this section we discuss some of the differences that one would encounter for fermions.

The chiral protection of fermion masses makes them plausible low energy remnants of the compactification of extra dimensions, also if the typical scale of compactification is much larger than their mass. For example, light fermions are common in some realizations of string theory~\cite{Halverson:2018xge}. One example that would have a collider phenomenology similar to the theories in Sect.~\ref{sec:models} is a model of $N$ Dirac fermions with a Yukawa interaction
\begin{equation} \label{eq:fermions}
\mathcal{L}_\psi = i \overline \psi_\alpha\gamma^\mu \partial_\mu \psi_\alpha - m_\alpha \overline\psi_\alpha \psi_\alpha
- y_{\alpha \beta}\overline \psi_\alpha \psi_\beta \phi+{\rm h.c.} .
\end{equation}
This choice is rather appealing from the point of view of coupling this sector to the SM since it gives us a symmetry argument to forbid the $l H\psi_\alpha$ vertex leaving us with the interaction
\begin{equation}
\mathcal{L}_{\psi H} = - \frac{\overline \psi_\alpha \psi_\beta |H|^2}{\Lambda} +{\rm h.c.} ,
\end{equation}
which gives the long decay chains discussed in Sect.~\ref{sec:pheno}. If we also take the mass of $\phi$ in a range that makes three-body decays within the dark sector ($\psi_\alpha \to \psi_\beta \phi^* \to \psi_\beta \psi_\gamma \psi_\delta$) relevant we can reproduce exactly the same structure that we had in the scalar models. 

However when considering the mass matrices of $N$ fermions one should be aware of some differences compared to the scalar case. We have to diagonalize $M^\dagger M$ to obtain the pole masses in the free theory, while the parameters in the Lagrangian form a linear mass matrix $M$. If we treat the the entries of $M$ as the fundamental parameters of the theory, drawn from a random distribution, the asymptotic form of the spectral density is different from what discussed for GOE and GUE.

Given a square $N\times N$ matrix $M$, with independent and identically-distributed entries drawn from a Gaussian distribution with zero mean and unitary variance the asymptotic distribution for the eigenvalues of $(M^\dagger M)/N$ is a particular case of the Mar\v{c}enko-Pastur density~\cite{MarcenkoPastur}
\begin{equation}
\rho_{\rm MP}(m)=\frac{1}{2\pi m}\sqrt{m (4 -m)} ,
\end{equation}
which denotes an accumulation of eigenvalues around zero. A diagrammatic derivation of this result can be found in~\cite{Lu:2014jua}. Note that we have used $m$ for the eigenvalues of $M^\dagger M$ that have the dimension of a squared mass. The Mar\v{c}enko-Pastur density can be easily generalized to the case of non-unitary variance. Assuming that all the entries of $M$ have zero mean and variance $\sigma$, the asymptotic distribution for the spectral density of $(M^\dagger M)/N$ is
\begin{equation}
\rho_{MP}(m, \sigma\neq0)=\frac{1}{2\pi \sigma^2 m}\sqrt{m (4 \sigma^2 - m)} .
\end{equation}
Also in this case turning on a non-zero mean $\mu$ equal for all the entries does not affect the spectral density except for the appearance of one large eigenvalue of $M$ that is $\mathcal{O}(N\mu)$. We leave a more detailed discussion to future work.

%**************** Section ***********************************
\section{Outlook}
\label{sec:outlook}
%*************************************************************

In this paper, motivated by the current null results at colliders, we have ignored some of the unspoken rules of BSM model building. We have asked what is the simplest scenario in which new particles are present at the weak scale, but still invisible at colliders, without attempting to answer the open questions that have driven the field. Even if our starting point was orthogonal to most traditional phenomenological studies, our ``kinematics of invisibility'' is realized in a number of scenarios that are well motivated theoretically.

More concretely, we have discussed the simple kinematical consequences of having a large number of new particles in a finite mass range. This situation can arise in many BSM scenarios and has characteristic phenomenological signatures that we have explored in Sect.~\ref{sec:pheno}. The main messages are that final states with multiple soft particles can be common and that the total production cross section can easily be small enough to motivate the high luminosity upgrade of the LHC. Often there is not enough total energy to pass global triggers such as those targeting high $H_T$ or missing $E_T$ signatures. This provides further motivation for the ongoing work aimed at lowering trigger thresholds.

The new sectors that we have studied are naturally described in terms of disordered mass matrices and couplings. This is a convenient phenomenological tool to parametrize our ignorance. It allows us to keep parameters that are supposed to be of the same order close to each other without giving up potentially interesting $\mathcal{O}(1)$ variations. The main phenomenological interest lies in accidental compressions near the bottom of the spectrum. We find that this situation is not uncommon and can considerably increase the number of soft final state particles in the event. Aside from this point, disordered mass matrices have an interesting structure that we have discussed in Sect.~\ref{sec:disorder} and expanded upon in the appendices. We have made an effort to rederive all relevant results in a language as close as possible to QFT. We hope that further explorations of disorder in model building will have interesting implications for the long standing questions in the field. Even if just at the level of intriguing coincidences we already find much more structure than we naively expected.

There are a number of new directions that this work suggests. Most of them are simple such as expanding the analysis of fermion models and of the combinatorial properties of long decay chains. However the one that we find most intriguing is the general exploration of large $N$ sectors and disorder in phenomenology, especially their aspects that we have not touched in this paper as the possibility of having a large number of metastable vacua and glassy phases.

In conclusion we have presented a simple phenomenon that motivates new explorations of hadron collider data, found connections with motivated BSM scenarios, and introduced some of the tools of large $N$ disordered models in a particle physics context.

\section*{Acknowledgements}
We would like to thank Y. Bai, T. Cohen, B. Dobrescu, H. D. Kim, and G. Villadoro for useful discussions.  RTD is supported by the U.S. Department of Energy under grant number DE-AC02-76SF00515. ML acknowledges support from the Institute for Advanced Study and from the Fermi Research Alliance, LLC, under Contract No. DE-AC02-07CH11359 with the U.S. Department of Energy, Office of Science, Office of High Energy Physics.  We thank the Galileo Galilei Institute for Theoretical Physics for the hospitality and the INFN for partial support during the completion of this work. This work was partially supported by a grant from the Simons Foundation (341344, LA).
 
%**************** Appendix ***********************************
\appendix
\section{The Vandermonde Determinant}
\label{app:vandermonde}
%*************************************************************

As discussed in Sect.~\ref{sec:disorder} the joint entry distribution for real symmetric $N \times N$ Gaussian-random matrices with variance $\sigma$ and mean equal to zero is
\begin{equation} \label{eq:rhoM}
 \rho[M] 
= \prod_{i=1}^N \frac{e^{-\frac{M_{ii}^2}{2 \sigma^2}}}{\sqrt{2\pi}\sigma}\prod_{i<j} \frac{e^{-\frac{M_{ij}^2}{\sigma^2}}}{\sqrt{\pi}\sigma} 
\propto e^{-\frac{1}{2 \sigma^2}{\rm Tr}[M^2]}.
\end{equation}
The joint eigenvalue distribution for the eigenvalues $m_i$ of the matrix $M$ is
\begin{equation} \label{eq:jointapp}
\hat{\rho}[m_1, \ldots, m_N] = \frac{1}{Z_{N}} e^{-\frac{1}{2\sigma^2}\sum_{i=1}^N m_i^2} \prod_{i>j} |m_i-m_j|,
\end{equation}
where $Z_{N}$ is a normalization factor.

The factor $|J|=\prod_{i>j} |m_i-m_j|$ is the Jacobian of the transformation between matrix entries and its eigenvalues and eigenvectors.  It is known as the Vandermonde determinant.  In this appendix we derive it in two ways.

The first is through a simple change of basis.  Starting from the joint entry distribution we find
\begin{equation} \label{eq:changeofbasis}
\rho[M] DM = \rho[M_D] |J(M_D)| Dm DU.
\end{equation}
To compute $J$, recall the definition of the metric on the space of symmetric matrices
\begin{equation} \label{eq:metric} \begin{aligned}
{\rm Tr}[dMdM] 
& = {\rm Tr}[d(OM_DO^T)d(OM_DO^T)] \\
& = {\rm Tr}[dm^2+[M_D, O^T dO]^2] \\
& = \sum_{i=1}^N dm_i^2+\sum_{i \neq j}(m_i-m_j)^2 (O^T dO)_{ij}^2 ,
\end{aligned} \end{equation}
where we have used $d(OO^T)=0$ which means $dO^T=- O^T dO O^T $.  This also shows that $O^T dO$ is an antisymmetric matrix.  We now have a metric tensor and we can use the square root of its determinant to obtain $J$.

Note that when we write $\rho[M]DM$ we are integrating only over the $N(N+1)/2$ independent variables in $M$.  Given the form of $\rho[M]$, integrating over the other matrix components would just give an overall constant absorbed by the normalization.  So we do not need all the components of the metric tensor defined by Eq.~\eqref{eq:metric} to compute $J$.  Combining Eq.~\eqref{eq:changeofbasis} and Eq.~\eqref{eq:metric} we obtain
\begin{equation}
|J|=\prod_{i> j}|m_i-m_j|\, .
\end{equation}
It is not hard to generalize these steps to the case of complex Hermitian matrices. 

The second way to derive $J$ is as a Fadeev-Popov determinant. The correlation functions of gauge-invariant operators $O_i(A)$ in a Yang-Mills theory
\begin{equation}
\frac{1}{Z}\int \mathcal{D} A e^{iS(A)} O_1(A) \cdots O_N(A),
\end{equation}
have the same structure as the expectation values of quantities that depend only on the eigenvalues of $M$
\begin{equation} \label{eq:matrixmodel}
\int DM \rho[M] O_1(m) \cdots O_N(m),
\end{equation}
if we identify $ \rho[M]$ with the action $e^{iS(A)}$.  In the matrix case a gauge transformation is a change of basis. From Eq.~\eqref{eq:rhoM} we see that an orthogonal transformation on $M$ does not change $\rho[M]$
\begin{equation} \label{eq:gauge}
 \rho[M] = \rho[OMO^T], 
\end{equation}
and leaves us in the space of symmetric matrices $(OMO^T)^T=OMO^T$. It is then easy to conclude that also in Eq.~\eqref{eq:matrixmodel} the action and integration measure are gauge-invariant. 

Following the Fadeev-Popov procedure we first define $\Delta(M_D)$ as
\begin{equation} \label{eq:FP}
1=\int Dm DU \delta [M-M_D] \Delta[M_D],
\end{equation}
where $\delta[M-M_D]$ is an $N(N+1)/2-$dimensional Dirac delta function for symmetric matrices $M$.  For complex Hermitian matrices it would be $N^2-$dimensional.

From Eq.~\eqref{eq:FP} we compute $\Delta[M_D]$ for an infinitesimal orthogonal transformation $O\approx 1_{N\times N}+\delta O$ which means $DU\approx d \delta O$.  We find
\begin{equation}\begin{aligned}
\Delta(M_D)
& = \frac{1}{\int Dm DU \delta[M-M_D]} \\
& \approx \frac{1}{\prod_{i>j}\int d\delta O_{ij} \delta[\delta O_{ij} D_{jj}+D_{ii}\delta O^T_{ij}]} \\ 
& \approx \frac{1}{\prod_{i> j} d \delta O_{ij}\delta\left[(m_i-m_j)\delta O_{ij}\right]} \\
& = \prod_{i> j}|m_i-m_j| .
\end{aligned}\end{equation}
So we have found that $\Delta(M_D)$ is precisely the Jacobian of the transformation. For a complex Hermitian $M$, $|J|=\prod_{i>j} |m_i-m_j|^2$ where the square derives from the double integration on the real and imaginary parts of the elements of the unitary transformation.

The last step to derive Eq.~\eqref{eq:jointapp} consists in multiplying Eq.~\eqref{eq:matrixmodel} by the identity in the form given in Eq.~\eqref{eq:FP} and note that $\int DU \delta[M-M_D]$ being the inverse of $\Delta(M_D)$ cancels the factor of $\prod_{i> j}|m_i-m_j|$ in the integration measure. The Jacobian is then restored by $\Delta(M_D)$.

%**************** Appendix ***********************************
\section{Moments of Wigner's Semicircle}
\label{app:moments}
%*************************************************************

Recall the Wigner semicircle distribution:
\begin{equation}
\rho_{\rm SC}(m)=\left\{\begin{array}{cc} 
\frac{1}{2\pi}\sqrt{4-m^2} & |m|<2, \\ 
0                          & |m|>2.\end{array}\right.
\end{equation}
The odd moments of this distribution vanish by symmetry.  Here we show that the even moments, $\mu_{2n}$, for integer $n$, are the Catalan numbers, $C_n$, where
\begin{equation}
C_n = \frac{1}{n+1} \binom{2n}{n}.
\end{equation}
We compute the integral in polar coordinates
\begin{equation} \label{eq:step2}
\mu_{2n} 
= \int^2_{-2} \rho_{\rm SC}(m) m^{2n} dm
= \frac{2^{2n+1}}{\pi} \int^{\pi/2}_{-\pi/2} \sin^{2n}\theta \cos^2\theta d\theta.
\end{equation}
Using a reduction formula
\begin{equation}
\int^{\pi/2}_{-\pi/2} \sin^n\theta d\theta = \frac{n-1}{n} \int^{\pi/2}_{-\pi/2} \sin^{n-2}\theta d\theta,
\end{equation}
the integral in Eq.~\eqref{eq:step2} can be simplified to an integral over a single power of sine
\begin{equation}
\mu_{2n} 
= \frac{2^{2n+1}}{\pi} \frac{1}{2(n+1)}\int^{\pi/2}_{-\pi/2} \sin^{2n}\theta d\theta\, .
\end{equation}
Finally using the identity
\begin{equation}
\int^\pi_{-\pi} \sin^{2n}\theta d\theta = \frac{\pi}{2^{2n-1}} \binom{2n}{n},
\end{equation}
one finds that
\begin{equation}
\mu_{2n} 
= \frac{1}{n+1} \binom{2n}{n}= C_n.
\end{equation}

%**************** Appendix ***********************************
\section{The Catalan Numbers and Planar Diagrams}
\label{app:catalan}
%*************************************************************

Here we present a partial derivation of the Wigner semicircle distribution using Feynman diagram techniques.  This derivation highlights the connection between Catalan numbers and planar diagrams.

Recall that the Catalan numbers are the even moments of the Wigner semicircle distribution
\begin{equation} \label{eq:evenmoments}
C_n = \int_{-2}^{2} \rho_{SC}(m) m^{2n} dm.
\end{equation}
Consider a Hermitian matrix $M$.\footnote{We show the Feynman rules for the GUE because they are slightly simpler than for the GOE.}  It has an associated Green's function
\begin{equation}
G_M(z) 
= \frac{1}{N} {\rm Tr}\frac{1}{z-M} 
= \frac{1}{N} \sum_{i=1}^N \frac{1}{z-m_i},
\end{equation}
where $m_i$ are the $N$ eigenvalues of $M$.  We are interested in the average over many realizations of $M$
\begin{equation} \label{eq:greensfunction}
G(z)
= \lim_{N\to \infty} \langle G_M(z) \rangle
= \int \frac{\rho(m)}{z-m} dm.
\end{equation}
The averaging merges the poles into a cut spanning the support of $\rho(m)$.  From the Green's function one can find the spectral density $\rho(m)$ via the identity
\begin{equation}
\rho(m) = -\frac{1}{\pi} \lim_{\epsilon\to0^+}{\rm Im} \; G(m + i\epsilon).
\end{equation}
We compute $G(z)$ using the Feynman diagram expansion developed in~\cite{Brezin:1993lnq} and discussed pedagogically in~\cite{Zee:2003mt}.  It is convenient to first introduce $G^i_j(z)$ as
\begin{equation} \label{eq:Gdef}
G^i_j(z) 
= \left\langle \left(\frac{1}{z-M}\right)^i_j \right\rangle 
= \delta^i_j G(z).
\end{equation}
We expand $G^i_j(z)$ to find
\begin{equation} \label{eq:greenexpansion}
G^i_j(z) = \sum_{n=0}^\infty \frac{1}{z^{2 n+1}} \langle (M^{2n})_{j}^i \rangle.
\end{equation}
Odd powers of $M$ vanish in the average.  The numerator of each term, explicitly, is
\begin{equation} \label{eq:path}
\langle (M^{2n})_{j}^i\rangle = \frac{1}{Z} \int dM e^{- \frac{N}{2}{\rm Tr}(M^2)} (M^{2n})_{j}^i,
\end{equation}
where $Z$ is a normalization factor.  Eq.~\eqref{eq:path} resembles a path integral for the matrix $M$.  Computing the full propagator requires evaluating these integrals which we can do using Feynman diagrams.

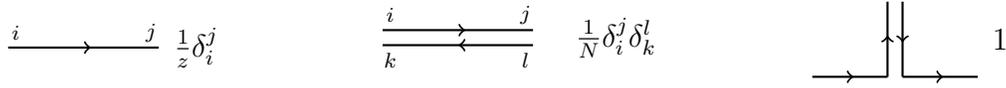
\begin{figure}[!h]\begin{center}
  \begin{tikzpicture}[scale=1.0]
  \draw[fermion,line width=0.8pt] (  0,   0) -- ( 2, 0);
  \node at                        (0.1, 0.2) {\scriptsize{$i$}};
  \node at                        (1.9, 0.2) {\scriptsize{$j$}};
  \node at                        (2.5,   0) {$\frac{1}{z} \delta_i^j$};
  \end{tikzpicture}
  \hspace{4em}
%%%%%%%%%%%%%%%%%%%%%%%%%%%%%%%%%%%%%%%%
  \begin{tikzpicture}[scale=1.0]
  \draw[fermion   ,line width=0.8pt] ( 0,  0.1) -- ( 2,  0.1);
  \draw[fermionbar,line width=0.8pt] ( 0, -0.1) -- ( 2, -0.1);
  \node at                           (0.1, 0.3) {\scriptsize{$i$}};
  \node at                           (1.9, 0.3) {\scriptsize{$j$}};
  \node at                           (0.1,-0.3) {\scriptsize{$k$}};
  \node at                           (1.9,-0.3) {\scriptsize{$l$}};
  \node at                           (3.1,   0) {$\frac{1}{N}\delta_i^j \delta_k^l$};
  \end{tikzpicture}
  \hspace{4em}
%%%%%%%%%%%%%%%%%%%%%%%%%%%%%%%%%%%%%%%%
  \begin{tikzpicture}[scale=1.0]
  \draw[fermion   ,line width=0.8pt] (-1.1,   0) -- (-0.1,  0);
  \draw[fermion   ,line width=0.8pt] (-0.1,   0) -- (-0.1,1.0);
  \draw[fermionbar,line width=0.8pt] ( 1.1,   0) -- ( 0.1,  0);
  \draw[fermionbar,line width=0.8pt] ( 0.1,   0) -- ( 0.1,1.0);
  \node at                           ( 1.4, 0.5) {$1$};
  \end{tikzpicture}
%%%%%%%%%%%%%%%%%%%%%%%%%%%%%%%%%%%%%%%%
  \captionsetup{width=0.87\linewidth}
  \caption{Feynman rules for random matrix theory to compute Eq.~\eqref{eq:path}.}
  \label{fig:feynmanrules}
%%%%%%%%%%%%%%%%%%%%%%%%%%%%%%%%%%%%%%%%
\end{center}\end{figure}

Using Feynman diagrams we can see that in the large $N$ limit planar diagrams dominate, similar to large $N$ QCD~\cite{tHooft:1973alw}.  The Feynman rules are shown in Fig.~\ref{fig:feynmanrules}.

%%%%%%%%%%%%%%%%%%%%%%%%%%%%%%%%%%%%%%%

\begin{figure}[!h]\begin{center}
  \begin{tikzpicture}[scale=1.0]
    \draw[fermion   ,line width=0.8pt] (  0,  0) arc (0:180:1);
    \draw[fermionbar,line width=0.8pt] (0.2,  0) arc (0:180:1.2);
    \draw[fermion   ,line width=0.8pt] ( 0.2,   0) -- ( 1.2,   0);
    \draw[fermion   ,line width=0.8pt] (-2.0,   0) -- (   0,   0);
    \draw[fermion   ,line width=0.8pt] (-3.2,   0) -- (-2.2,   0);
    \node at                       (-3.1, 0.3) {\scriptsize{$i$}};
    \node at                       ( 1.1, 0.3) {\scriptsize{$j$}};
  \end{tikzpicture}
%%%%%%%%%%%%%%%%%%%%%%%%%%%%%%%%%%%%%%%
  \captionsetup{width=0.87\linewidth}
  \caption{The $n=1$ term from Eq.~\eqref{eq:Gdef} in the computation of $G^i_j(z)$.}
  \label{fig:feynman-n2}
\end{center}\end{figure}
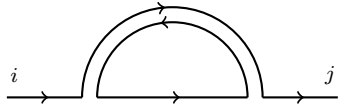

The $n=1$ term is shown in Fig.~\ref{fig:feynman-n2}.  Given that there is no integral over space or time, each diagram contributes a pure number to $\langle (M^{2n})_{j}^i\rangle$.  Closed loops correspond to factors of $N$ from tracing so that non-planar diagrams are suppressed by powers of $1/N$.

%%%%%%%%%%%%%%%%%%%%%%%%%%%%%%%%%%%%%%%

By inspecting diagrams, such as the one in Fig.~\ref{fig:feynman-n2}, we can conclude that there are as many planar diagrams with $n$ vertices as there are non-crossing partitions of a lattice with $n$ sites.  The definition of a non-crossing partition is precisely that if one puts the $n$ points of a lattice on a circle and connects each point with the next member of its part by an internal path (in cyclic order), the paths do not cross.  The Catalan numbers count, among other things, non-crossing partitions~\cite{Anderson2010}.  Therefore
\begin{equation}
G_j^i(z) = \delta_j^i \sum_{n=0}^\infty \frac{1}{z^{2 n+1}}C_n.
\end{equation}
With Eq.~\eqref{eq:greenexpansion} this implies
\begin{equation}
C_n = \int \rho(m) m^{2n} dm,
\end{equation}
which was shown explicitly in App.~\ref{app:moments}. 

\bibliographystyle{utphys}
\bibliography{refs}
\end{document}